\documentclass{article}
\title{The extended next release problem.\\ 
Generic Formulation of the Requirements Selection Problem\footnotemark }
\author{Isabel del~\'Aguila \\
	Univ. of Almer\'ia, Spain  \\
	\and 
	Jos\'e del~Sagrado \\
    	 Univ. of Almer\'ia, Spain  \\
    	\and 
	Alfonso Bosch \\
	Univ. of Almer\'ia, Spain  \\
	}

\date{}

\usepackage[a4paper, left=2cm, right=2cm, top=2cm, bottom=2cm]{geometry}
\usepackage{amsmath}
\usepackage{amsfonts}
\usepackage{graphicx}
\usepackage{rotating}
\usepackage{pdflscape}
\usepackage{tabulary} 
\usepackage{footnote}
\usepackage[square,numbers]{natbib}

\begin{document}





\maketitle

\footnotetext{Cite the short version of this work as: Del Águila, I. M., Del Sagrado, J., \& Bosch, A. (2024, October). Generic Formulation of the Requirements Selection Problem. In 2024 10th International Conference on Optimization and Applications (ICOA) (pp. 1-6). IEEE.}

\begin{abstract}
 Due to the limited amount of resources available for the next release of the current  product under development 
 not all stakeholders requests can be included in the next product to deliver. This optimization problem, known as the \emph{Next Release Problem} (\ensuremath{\mathsf{NRP}}), has customers satisfaction and development costs as the basic optimization objectives, and has been the subject of many research works. However, there are additional issues that deserve to be considered and included in the definition of the \ensuremath{\mathsf{NRP}}, such as supplementary optimization objectives including the elicited properties about the requirements,
 or the analysis of the non-dominated solution sets found   to decide which solution is preferred. This paper presents a generic formulation for this problem that allows the management of the currently agreed properties and relationships between requirements. It provides an open formulation that is capable of growing as the scope of the \ensuremath{\mathsf{NRP}} grows.
 We specify how our proposal can be used in software product projects when requirements selection has to be performed. We also describe how our formulation has been instantiated to cover previous solving approaches to this problem, using six case studies to demonstrate the successful customization of the generic formulation.

\textbf{Keywords}: Next Release Problem, Search Based Software Engineering, Requirements Engineering, Optimization Problems

\end{abstract}

.

\section{Motivation}

Requirements selection has been on the spotlight of many research works and the software industry. The proposed techniques exploit different information associated with the requirements. Some proposals use only one requirements' attribute, others a combination of them. Besides, both  stakeholders' and developers' sides could be considered inside the selection. The range of suitable techniques to be applied is broad and diverse~\cite{Berander2005,svahnberg2010,achimugu2014,thakurta2017,zhang2018,hujainah2018}. However, both a deep knowledge of the domain and a skillful quantification capability have to be involved to succeed at using most of these techniques~\cite{SWEBOK2014}.

This selection problem has also been formulated as an optimization problem called Next Release Problem (\ensuremath{\mathsf{NRP}})~\cite{bagnall2001,zhang2007}. The goal is to find the ideal set of requirements or customers that balance stakeholders' satisfaction within a set of fixed constraints, such as resource or effort constraints. There are many solving techniques that have been applied in order to find this set of requirements including hill climbing~\cite{bagnall2001}, simulated annealing ~\cite{bagnall2001,sagrado2011}, algorithms based on genetic inspirations~\cite{greer2004,finkelstein2009, durillo2011}, nature inspired optimization~\cite{sagrado2010,sagrado2015,alrezaamiri2020,piroz2021}, stochastic optimization ~\cite{sagrado2023}, linear programming~\cite{bagnall2001,van2005,dominguez2019}, or clustering approaches~\cite{sagrado2021,HUJAINAH2021}, among others.

Being a problem with wide coverage and  well established, some issues have not been addressed in the usual formulation (i.e., restraining the applicability of the solutions obtained) or have been only partially addressed, such as the need to manage the heterogeneity of stakeholders that can lead to tensions in the project~\cite{zhang2011}, the asymmetric satisfaction behavior~\cite{nayebi2018}, or the coverage of value interactions~\cite{mougouei2021}. Therefore, there is neither a unified nor a general \ensuremath{\mathsf{NRP}} formulation that can encompass all of these current alternatives and allowing for its extension in the future.

This work proposes an extended version of the next release problem (\ensuremath{\mathsf{extNRP}}) defining an open generic formulation that can grow and be adapted along with
the software engineering methodologies evolution in the forthcoming  future. The extended version, \ensuremath{\mathsf{extNRP}}, will allow the use of a no limited set of optimization objectives and constraints based on the different requirements attributes and interactions elicited for the project at hand.

On the one hand, our proposal is useful for researchers, academic or otherwise, in the problem of requirements selection, providing them with a general framework that facilitates, not only the embedding of previous research, but also the development of new proposals using, instantiating, or improving upon this extended version of \ensuremath{\mathsf{extNRP}}. On the other hand, for those readers looking for a practical method of applying search algorithms to the problem of requirements selection within their organisation, our extended definition allows the unrestricted use of optimisation objectives. This versatility facilitates the inclusion of any specific data collected in their projects to automatically apply search algorithms to obtain reliable solutions to their requirements selection problems. The concise case study in Section~\ref{subsec:didactic_ex} and the examples of use described in Section~\ref{sec:instances} are intended to illustrate and facilitate the aspects of practical use.

The depiction of our formulation follows the research questions outlined below: 

\begin{itemize}

\item \textbf{\emph{RQ1}}  \emph{Articulateness.} 
Is it possible to define an extended formulation for the next release problem (\ensuremath{\mathsf{extNRP}}) that allows the inclusion of an unlimited number of optimization objectives and constraints based on the requirements attributes and interactions?

\item  \textbf{\emph{RQ2}}  \emph{Competence.}  Does the use of  \ensuremath{\mathsf{extNRP}} facilitate the identification of the ideal set of requirements to be included in the next software version?

\item  \textbf{\emph{RQ3}} \emph{Coverage.} Can  the previous \ensuremath{\mathsf{NRP}} formulations existing in the literature be covered by this new formulation, \ensuremath{\mathsf{extNRP}}?

\end{itemize}

 The structure of the rest of the paper begins in Section~\ref{sec:related-work} with an overview of the most important \ensuremath{\mathsf{NRP}} previous works. Section~\ref{sec:data} describes the different information available about requirements and how to manage it in the problem.  After presenting the current optimization approach to the \ensuremath{\mathsf{NRP}} in Section~\ref{sec:staged-approach}, we provide a new open formulation under the form of a generic optimization problem, \ensuremath{\mathsf{extNRP}}, in Section~\ref{sec:formal-definition}. The use of the proposed formulation, in particular how to analyze the results of the optimization algorithms approaching the problem to the product world is
shown in  Section~\ref{sec:analysis}. Section~\ref{sec:instances} instantiates the \ensuremath{\mathsf{extNRP}}) for six previous cases of \ensuremath{\mathsf{NRP}}.  Finally, Section~\ref{sec:conclusions} concludes the paper.

\section{Related works}
\label{sec:related-work}

Putting it down in words, the next release problem can be defined as \emph{how to select a set of features or requirements to be covered by the application (or its the new release) that is being developed when they face contradictory goals}. One of the first works about \ensuremath{\mathsf{NRP}} were done by Karlsson~\cite{karlsson1996} in the 1990s. It's was not until the next decade that the term Next Release Problem \ensuremath{\mathsf{NRP}} was coined by the works of Bagnall et al.~\cite{bagnall2001} and Van den Akker et al.~\cite{van2005}. These two initial formulations intertwined offering to the software engineering community a solid set of different strategies that have been in development for the last twenty years to solve this problem  using a search based approach~\cite{pitangueira2015,hujainah2018}.

Since then,  several optimization schemes (single-objective, bi-objective, multi-objective) have been deeply studied. The original single-objective formulation~\cite{bagnall2001,van2005} deals with the search of the set of requirements or customers that maximizes the stakeholders' satisfaction, without violating the restriction defined for the development effort bound ($B$). That is, the implementation cost cannot exceed the available resources. According to this formulation, the \ensuremath{\mathsf{NRP}} is a single objective optimization problem based on the satisfaction concept~\cite{jung1998}. The bi-objective formulation of the problem~\cite{zhang2007} was obtained by incorporating cost/effort as the second objective to be optimized, giving solutions in the Pareto optimal sense. Other papers use more than two objectives~\cite{ramirez2019}. Some of them include as objectives the number of requirements and the fairness~\cite{finkelstein2009}, one objective for each customer considering their coverage~\cite{zhang2011}, dissatisfaction~\cite{nayebi2018, sangwan2020}, coverage and fairness~\cite{geng2018}, urgency~\cite{agarwal2014}, or risks~\cite{pitangueira2017}.

In the case of the single objective formulation the main solving techniques applied are  hill climbing~\cite{bagnall2001}, simulated annealing ~\cite{bagnall2001,sagrado2011}, integer linear programming~\cite{bagnall2001,van2005}, genetic algorithms~\cite{greer2004}, ant colony optimization~\cite{sagrado2010,sagrado2011}, or the approximate backbone based multilevel algorithm~\cite{Xuan2012}. Further, when several objectives are considered other algorithms have been applied, such as  algorithms based on genetic inspiration~\cite{zhang2007,durillo2011}, the application of differential evolution (i.e., evolutionary algorithms)~\cite{chaves2015}, the use of nature inspired optimization  (e.g., ant colony~\cite{sagrado2015}, bee colony~\cite{alrezaamiri2020}, whale and grey wolf optimization~\cite{ghasemi2021} or, algae algorithm~\cite{piroz2021}),  linear programming~\cite{dominguez2019}, clustering approaches~\cite{sagrado2021,HUJAINAH2021} or even,  exact methods to find the whole Pareto front~\cite{DONG2022}. Other studies go one step beyond  looking for more complex techniques applying hyper-heuristics as a search methodology that  automatize the process of selecting or combining simpler heuristics~\cite{zhang2018}, or proposing methods that are able to automatically learn from programs that solve optimization problems (Virtual Savant)~\cite{massobrio2021}.
 
Since requirements often have dependencies or interactions that force some requirements to be implemented before others or to be dropped due to mutual exclusions, another point addressed  in the literature is the management of the requirements interactions. First works that defined the different types of interactions~\cite{karlsson1996, Carlshamre2001} were followed by the use of graphs to represent implication dependencies~\cite{ngo2004} which was later extended to include all the functional dependencies between requirements~\cite{sagrado2015}. 
These functional interactions are problem constraints, being precedence/implication ones those that limit the search space~\cite{sagrado2011}. Other kinds of interactions, which are related to the increment/decrement of some attributes values assigned to the requirements (i.e., revenue-based and cost-based~\cite{Carlshamre2001}), have been addressed  using additional elements such as preference matrices~\cite{mougouei2021} or pre- and post-processing actions~\cite{sagrado2015}.

Nonetheless, the research on these interactions reveals that few of them capture valuable relationships between requirements~\cite{zhang2014}. Approaches to embed interactions in NRP vary from linear programming ~\cite{li2010,veerapen2015}, the use of a layered metamodel~\cite{aydemir2018}, graphs and matrices ~\cite{sagrado2015,sangwan2020,mougouei2021}, or the use of interactions to define an underlying probabilistic graphical model ~\cite{sagrado2023}. For instance, some of them (e.g., combination) can be addressed prior to the execution of the optimization algorithm, while others (i.e., implication, revenue-based, and cost-based) need to be embedded in the optimization algorithm~\cite{sagrado2015}.

Finally, the analysis of the set of solutions (i.e., the Pareto front) obtained for a \ensuremath{\mathsf{NRP}}, when more than one objective is included in its formulation, has been addressed in several works,  because the \ensuremath{\mathsf{NRP}} is not finished when the Pareto front is available~\cite{ferreira2007}, as the solutions have not yet been compared. Some authors suggest explaining the results using quality indicators closer to the problem domain~\cite{aguila2016} or using tools to support optimization~\cite{zhang2020,aguila2016}.

In summary, \ensuremath{\mathsf{NRP}} is an open research area, as it involves the use of different optimization schemes, solving techniques, management of requirement interactions, and also there is a need of analyzing the set of solutions obtained. None of the current formulations covers all of these points. The new \ensuremath{\mathsf{extNRP}} is intended to be a general framework that accommodates, in a single formulation, all aspects that have been considered so far, and even be open enough to incorporate new aspects that may appear in the future.  The next section will therefore present and review the requirements selection issues that influence the extension of \ensuremath{\mathsf{NRP}}, specifically the relevant requirements data that we need to consider when selecting requirements in a software engineering project.

\section{Requirements data} \label{sec:data}

To build a useful technological system, we need to know its requirements. This means that eliciting, documenting and validating stakeholder requirements are fundamental activities. Requirements and their descriptions are the starting point for delivering a new version of the product with the additional capabilities defined by new requirements.

Let $\mathbf{Re} = \{r_1, r_2, \cdots, r_q\}$ be the set of requirements considered in a project. Requirements express the needs to be included in a software product, and collect the functions that the software will perform. The requirements features that according to the 
 fundamental literature~\cite{babok2015,SWEBOK2014,ireb} have been agreed by Requirements Engineering practitioners and experts are: Value of benefit, Penalty, Cost, Risk,  Time sensitivity, Stability/Volatility, Regulatory or policy compliance, and Dependencies/interactions. These properties characterize them and allow us to define the objective functions when prioritization is needed~\cite{babok2015,Berander2005}.

\subsection{Requirements interactions}
The first thing to note is that interactions between requirements can alter the value of other attributes. 
There are many models to describe  \textbf{\emph{requirements interactions or dependencies}}, nonetheless, there is no unified set of types, nor clear definitions to facilitate their applicability. 
Some categories are difficult to understand or have duplicated definitions, or even being properly defined, no project where to apply them can be found~\cite{zhang2014}. Besides, not all of them represent relationships between requirements, such as \emph{traceability interactions} that mainly are connections between requirements and the rest of the project artifacts. 

\emph{Structural interactions} represent semantic relationships between requirements. In this group, \emph{refine}, \emph{similar}, and \emph{restrict} interactions can be found~\cite{babok2015}. The first, \emph{refine}, represents hierarchies, or levels between requirements, and the second one, \emph{similar}, identifies groups of requirements on the same concept, such as CRUD (create, replace, update, delete). For non-functional requirements, a \emph{restrict} interaction limits all the functional ones that are related to. Other types of interactions are \emph{satisfies}, expressing that if the target requirement is reached, the source object is also satisfied, 
or \emph{evolve into} that represents a connection between the new version of a requirement and the previous one \cite{zhang2014}.

A particular group of structural interactions are the so-called \emph{functional interactions}, they impose a specific implementation order, downsizing the \ensuremath{\mathsf{NRP}} feasible solution set~\cite{sagrado2011}.  \emph{Implication}, \emph{combination} and \emph{exclusion} dependencies are the functional interactions that have usually been studied in the NRP literature~\cite{bagnall2001,sagrado2015,veerapen2015,geng2018}.

Additionally, original versions of \ensuremath{\mathsf{NRP}} include two value-based interactions: \textit{revenue-based} and \textit{cost-based}~\cite{Carlshamre2001}. These interactions express situations where the inclusion in the same version of a set of requirements implies an increment or decrement either in the effort (i.e., cost-based) or in the satisfaction (i.e., revenue-based). This kind of interactions had been modeled using graphs~\cite{mougouei2021} or matrices~\cite{sagrado2015,li2010}. 

However, it should be noted that a requirement can be characterized not only by two (i.e. effort and satisfaction), but by $k$ properties that give rise to $k$ different types of value interactions.
This generalization results in a new category, which we will call \emph{value interactions}, to be considered for all attributes that characterize requirements.

Consequently, the initial graph based approach to interactions representation proposed in Sagrado et al.~\cite{sagrado2015} has to be adapted in order to incorporate \emph{refine} and \emph{value}-based interactions. More formally, let $\mathbf{Re} = \{r_1, r_2, \cdots, r_q \}$ be the set of requirements considered in a project. A subset $\mathbf{N} \subseteq \mathbf{Re} \times \mathbf{Re}$ defines not valued-based interactions, whereas $\mathbf{N}^\prime \subset \mathbf{Re} \times \mathbb{R}$ defines valued-based interactions. The subsets representing interactions are:

\begin{itemize}
\item $\mathbf H=\{ (r_i, r_j) \mid r_i$  \emph{is refined by} $r_j \}$, represents that requirement $r_i$ has been broken down during elicitation being $r_j$ one of its refinement in the requirements hierarchy. 

 \item $\mathbf I=\{ (r_i, r_j)\mid r_i$ \emph{implies}  $r_j \}$, it models that a requirement $r_i$ needs to be implemented before the requirement $r_j$. 

\item $\mathbf J=\{ (r_i, r_j) \mid r_i$ \emph{combined with} $r_j \}$ both requirements have to be developed in the same iteration.

\item $\mathbf{Ex} = \{ (r_i, r_j) \mid r_i$ \emph{excludes} $r_j \}$. Each pair $(r_i, r_j)$ reveals that both requirements are incompatible. That is, they could not be developed in the same product. 

\item $\mathbf{Y_a}= \{(\mathbf{U}, qty) \mid \mathbf{U} \subseteq \mathbf {Re}, ~ qty \in \mathbb{R} \text{ is a constant value} \}$, when all $r_i \in \mathbf{U}$ are considered in the iteration,  then for each requirement $r_i \in \mathbf{U}$  the value of the attribute $\mathbf{a}$ has to be modified according to the quantity indicated by $qty$.
\end{itemize}

\subsection{Requirements attributes}
\label{sec:attributes}

The information associated to requirements is given in terms of a set of agreed attributes~\cite{babok2015,SWEBOK2014,ireb}: \emph{Value of benefit, Penalty, Cost, Risk,  Time sensitivity, Stability/Volatility, Regulatory or policy compliance}. Let $\mathbf {R} =\{r_1, r_2, \ldots, r_n\}$, $\mathbf {R} \subseteq \mathbf {Re}$, be the set of requirements that are being considered in a selection problem. The set $\mathbf{At} =\{a_1, a_2, \ldots, a_k\}$ represents the list of properties or attributes of each requirement. For each attribute $a_j \in \mathbf{At}$, there is a set $\mathbf{V_{a_j}} = \{ (r_i, v_{a_j}) \vert r_i \in \mathbf{R}, v_{a_j} \in \mathbb{R} \}$, where each pair $(r_i, v_{a_j})$ indicates the value $v_{a_j}$ assigned to the requirement $r_i$ according to the attribute $a_j$.

The \textbf{\emph{Value of benefit/Value}} is the advantage or satisfaction that accrues to stakeholders as a result of a requirement implementation. The \textbf{\emph{Penalty}} represents the consequences that result from not implementing a given requirement. Because of the asymmetric behavior of value and penalty, both must be modeled separately~\cite{nayebi2018}, following also the pattern defined for the classical prioritization method, the Kano method~\cite{babok2015}. Stakeholders are responsible for setting the value and penalty of the requirements. Let $\mathbf{C} =\{c_1, c_2, \ldots, c_m\}$ be the set of stakeholders/customers involved in a project and 
$\mathbf{W} = \{w_1, w_2, \ldots, w_m\}$ be the set of weights representing the importance of the stakeholders. Each of these stakeholders can assign a value and a penalty to each of the requirements they are interested in. Therefore, there is no one-to-one cardinality between sources and attribute values. We call this type of requirements attributes \emph{multivalued}. It is common in the literature to calculate the value and penalty as an aggregation or combination of the values given by the stakeholders by means of a weighted sum. Requirements satisfaction and dissatisfaction represent the aggregation of value and penalty, respectively~\cite{nayebi2018, sangwan2020}.

Thus, for a given requirement $r_j \in \mathbf{R}$ its satisfaction, $s_j$, is defined as
\begin{equation} \label{eq:req_sat}
    s_j = \sum_{i=1}^{m} w_i * v_{ij},
\end{equation}

\noindent where $v_{ij}$ is the value that the stakeholder $c_i \in \mathbf{C}$ assigns to  requirement $r_j$. In this way, we obtain the set $\mathbf{S} = \{s_1, s_2, \ldots, s_n\}$ of satisfaction values needed to define $\mathbf{V_{a_{value}}}$.

Similarly, the dissatisfaction~\cite{nayebi2018} of a requirement $r_j$ is computed as 
\begin{equation} \label{eq:req_dis}
    d_j = \sum_{i=1}^{m}w_i * p_{ij}, 
\end{equation}

\noindent where $p_{ij}$ is the penalty that stakeholder $c_i$ assigns to the requirement $r_j$. And we obtain the set $\mathbf{D} = \{d_1, d_2, \ldots, d_n\}$ of dissatisfaction values needed to define $\mathbf{V_{a_{penalty}}}$.

From the set of values $\mathbf{S}$ and $\mathbf{D}$ of these requirements attributes, we can compute the satisfaction of a set of requirements $\mathbf{U} \subseteq \mathbf{R}$ as~\cite{bagnall2001,van2005}

\begin{equation}
\label{eq:set_sat}
\text{sat} (\mathbf{U}) = \sum_{r_j \in \mathbf{U}} s_j,
\end{equation}

\noindent and its dissatisfaction as~\cite{nayebi2018}

\begin{equation}
\label{eq:set_dis}
\text{dis} (\mathbf{U}) = \sum_{r_j \in \mathbf{U}}d_j.
\end{equation}

It should be pointed out that weighted aggregation is an option that can be used to calculate the value of any multivalued attribute that needs weights to determine the importance of its sources. For instance, if risks were defined by multiple developers, their years of experience would be used as weights for risks aggregation. However, alternative operators, such as maximum or minimum, can be used. But, in case the values of a multivalued attribute have been agreed or estimated directly, no calculation is necessary.

The \textbf{\emph{Cost/Effort}} of software development is often related to the number of staff hours. The requirement cost is expressed in units of effort (i.e., person-hours) and is an aggregation of the cost of the resources needed for its implementation: Elicitation effort, Design effort, Coding effort, and Testing effort. These effort categories or roles are defined according to the development methodology used in the project. For instance, if the Scrum framework is used, user story points will be the only cost to be considered for each requirement/user story.

Formally, let $\mathbf{R} = \{ r_1, r_2, \cdots, r_n \}$ be the set of requirements considered in a selection problem. The effort of a requirement $r_j\in \mathbf{R}$, represented by $e_j$, is defined as
\begin{equation}\label{eq:req_eff_categories}
e_j = \sum_{i \in \text{effort category}} ec_{ij}
\end{equation}
 \noindent where $ec_{ij}$ represents the effort value for requirement $r_j  $ in category $i$ defined according to the project methodology.
 Thus, we obtain the set $\mathbf{E} = \{e_1, e_2, \ldots, e_n\}$ of effort values needed to define $\mathbf{V_{a_{effort}}}$. As with the previous attributes, effort is a multivalued attribute and 
 its values can also be estimated directly. 
 
 From the set of values $\mathbf{E}$  we can compute the effort of a set of requirements $\mathbf{U} \subseteq \mathbf{R}$ as~\cite{zhang2007}

\begin{equation}
\label{eq:set_eff}
\text{eff} (\mathbf{U}) = \sum_{r_j \in \mathbf{U}} e_j.
\end{equation}

Some variations to the cost, such as \textbf{\emph{price}} (which is related to monetary resources) or \textbf{\emph{time}} (lead time is influenced by  other factors such as degree of parallelism, training needs, need to develop support infrastructure, complete industry standards, etc.) can be managed, as well. Thus, in a similar way, the price of a set $\mathbf{U} \subseteq \mathbf{R}$ of requirements is defined as

\begin{equation}
\label{eq:set_pri}
\text{pri} (\mathbf{U}) = \sum_{r_j \in \mathbf{U}} pr_j.
\end{equation}

\noindent where  $pr_j$  is  the development cost in monetary units of the requirement $j$.

A \textbf{\emph{Risk}} is an uncertain event or condition about a requirement. If it happens, the requirement could lose some or all of its potential benefit. However, there are a number of non-disjoint categories to which a risk may be assigned to, such as risks related to users, requirements, project complexity, planning/control, scheduling, quality, team, or organizational environment. This fact makes it neither easy, nor feasible to assign risks, and therefore a risk level, to individual requirements, and even more difficult to assign risks to the whole set of requirements. If a risk level is available for the requirements, riskier requirements should be left for forthcoming versions when agile methods are used, because backlog grooming could help to mitigate or adjust risky elements~\cite{popli2014}. In this case, it is preferred to minimize risks. Nevertheless, for predictive methodologies (i.e., cascade framework), the riskier requirements should be included as soon as possible, to be sure that the fails appear at the very beginning when the investment is not big~\cite{pitangueira2017}. 
Here, the preference is to maximize risks. Previous requirements selection works define the risk of a set $\mathbf{U} \subseteq \mathbf{R}$ of requirements using an multivalued aggregated approach as~\cite{du2014,pitangueira2017}
\begin{equation}
\label{eq:set_risk}
\text{risk} (\mathbf{U}) = \sum_{r_j \in \mathbf{U}} rk_j.
\end{equation}

\noindent where   $rk_j$ represents the risk of the requirement $r_j\in \mathbf{U}$

Some requirements have an expiration date. These requirements are \textbf{\emph{time sensitive}}, and they will lose value if they are not implemented as soon as possible~\cite{du2014}. Market conditions can be included as time sensitivity attributes. 
This property has been also called \textbf{\emph{urgency}}~\cite{agarwal2014}, and can be considered as a multivalued attribute. 
The urgency of a requirement $r_j \in \mathbf{R}$, represented by $n_j$, is defined as

\begin{equation}\label{eq:req_urg}
n_j = \sum_{i=1}^m  w_i * u_{ij}
\end{equation}

 \noindent where $u_{ij}$ is the urgency value estimated by stakeholder $c_i \in \mathbf{C}$ for requirement $r_j$, and $w_i \in \mathbf{W}$ is the weight (i.e., importance) of stakeholder $c_i$.
 Thus, we obtain the set $\mathbf{N} = \{n_1, n_2, \ldots, n_n\}$ of urgency values needed to define $\mathbf{V_{a_{urgency}}}$. 

From this set $\mathbf{N}$ of urgency values of the requirements, the \emph{time sensitivity} for a set of requirements $\mathbf{U} \subseteq \mathbf{R}$ is defined as
 
\begin{equation}
\label{eq:set_urg}
\text{tim}(\mathbf{ U}) = \displaystyle\sum_{r_j \in \mathbf{U }} n_j.
\end{equation}

A not stable requirement may be a bad candidate for the next version because further analysis might be needed, or stakeholders might not have reached an agreement yet. Were it included, some rework would be needed, wasting effort. While changes are common, because of the market, laws, or business policies, they cost money and create troubles, so a not stable requirement should have low priority. \textbf{\emph{Instability}} property is a multivalued attribute that represents how unstable a requirement. We define the \emph{instability} of a requirement $r_j \in \mathbf{R}$ in a selection problem, represented by $ins_j$, as

\begin{equation} \label{eq:req_ins}
\text{ins}_j = \sum_{i=1}^m  w_i * vl_{ij}
\end{equation}

 \noindent where $vl_{ij}$ is the volatility value that stakeholder (or developer) $c_i \in \mathbf{C}$ assigns to requirement $r_j$, and $w_i \in \mathbf{W}$ is the weight (i.e., importance) of stakeholder $c_i$. Thus, we obtain the set $\mathbf{INS} = \{ins_1, ins_2, \ldots, ins_n\}$ of instability values needed to define $\mathbf{V_{a_{instability}}}$. Then, we can define the instability of a set $\mathbf{U} \subseteq \mathbf{R}$ of requirements as

\begin{equation}
\label{eq:set_ins}
\text{ins}(\mathbf{ U}) = \displaystyle\sum_{r_j \in \mathbf{U }} ins_j.
\end{equation}

\textbf{\emph{Regulatory}} or \textbf{\emph{policy compliance requirements}} are mandatory and have to be included in the final product because they are stipulated by laws or by the business policies defined for the involved companies. Although these requirements affect the resources bounds for the \ensuremath{\mathsf{NRP}}, they should be out of the problem alternatives because their inclusion is compulsory. Therefore, the set of regulatory requirements $\mathbf{P} \subset \mathbf{Re}$ have to be removed from the set of requirements considered in the selection problem (i.e., $\mathbf{R}= \mathbf{Re} - \mathbf{P}$) and resources bounds have to be adjusted accordingly (e.g., the effort bound is adjusted by $B'= B-\text{eff}(\mathbf{P})$).

\subsection{Project conditions affecting the requirement data}

In a real project, there are certain aspects that need to be managed when selecting requirements because they affect and limit the valid candidate solutions; in particular, requirement interactions have a high impact on this problem. 

Let's look at how to deal with changes in attribute values and relationships between requirements when interactions 
occur.

The \emph{refine} interactions, $\mathbf H =\{ (r_i, r_j) \mid r_i$  \emph{is refined by} $r_j \}$, force us to rearrange the set of requirements to be managed based on the availability of attribute values. Consider requirement $r_i$, from the refinement interactions we define the set $\mathbf R^i = \{ r_k \vert (r_i, r_k) \in \mathbf H \}$ containing all requirements that refine it. A selection problem can include $\{ r_i \}$ or $\mathbf{R}^{i}$, but not both of them. Which one depends on at what level in the requirements' hierarchy are the attributes available at. That is, whether the value of the attributes is known for $\{ r_i \}$ or for its refinement $\mathbf{R}^{i}$. In the first case, the set $\mathbf{R}$ of requirements considered in the selection problem will satisfy $\{ r_i \} \subset \mathbf{R}$ and $\mathbf{R}^{i} \cap \mathbf{R} = \emptyset $, while in the second the opposite, $\{ r_i \} \cap  \mathbf{R} = \emptyset$ and $\mathbf{R}^{i} \subset \mathbf{R}$, is true.

Attribute values can be defined at any level in the hierarchy established by $\mathbf H$, but it does not always have to be the same level for all attributes, so it is necessary to define some propagation mechanism. Some authors propose value propagation to solve this question~\cite{lim2013}. However, this is a rather simplistic approach because not always a value such as the effort has to be uniformly distributed over the requirements at a lower level. Some refined requirements might require less effort than others. And conversely, for higher level requirements, an aggregation of the lower values, computed as a sum, is not necessarily the correct value, although it is commonly used in the requirements selection literature.  Alternative operators, such as maximum or minimum, can be used. Therefore, not all requirements have to be at the same level in the hierarchy.

\emph{Combination} interactions also  affect  the problem. Each pair ($r_i, r_j) \in \mathbf{J}$ generates a new requirement, $r_{i+j}$, which replaces the original two, $r_i$ and $r_j$. 
The value of each attribute for this new requirement is obtained by aggregating the values of the replaced requirements. Besides, the \emph{implication} interactions set, $\mathbf{I}$, should also be adjusted.   Any pair in $\mathbf{I}$ involving $r_i$ or $r_j$ must be modified by replacing them with $r_{i+j}$.

\emph{Value} interactions affect the value of  the requirement  attributes and thus the value of the functions calculated from them. Let be $\mathbf{Y_{a_i}}$ be the value interactions defined for attribute $a_i \in \mathbf{At}$. Elements in $\mathbf{Y_{a_i}}$ are pairs $(\mathbf{R_{a_i}}, y)$, where $\mathbf{R_{a_i}} \subset \mathbf{R}$ is a subset of requirements involved in the interaction, which causes the value of the attribute $a_i$ to change by the amount specified by $y \in \mathbb{R}$. Then, the total adjust induced by the value interactions $\mathbf{Y_{a_i}}$ to a set of requirements $\mathbf{U} \subseteq \mathbf{R}$ can be defined as
\begin{equation}\label{eq:adjust}
    \text{Adjust} (\mathbf{U}, \mathbf{Y_{a_i}}) = \sum_{\mathbf{R_{a_i}} \subset \mathbf{U}} y.
\end{equation}
This quantity represents the variation in any function $f_{a_i}$ associated with attribute $a_i$ when it is applied to a subset $\mathbf{U} \subseteq \mathbf{R}$ of requirements considered in the selection problem, and has to be included in its calculation as
\begin{equation}\label{eq:objective_fun}
    f_{a_i}(\mathbf{U}) = f_{a_i}(\mathbf{U}) + \text{Adjust} (\mathbf{U}, \mathbf{Y_{a_i}}).
\end{equation}

The value interactions commonly used in the \ensuremath{\mathsf{NRP}}  literature  are based on \emph{revenue} and \emph{costs}, and refer to the attributes of \emph{satisfaction} and \emph{effort}, respectively. Both, revenue-based interactions, $\mathbf{Y_{s}}$, and cost-based interactions $\mathbf{Y_{e}}$, are considered binary relationships involving two requirements~\cite{Carlshamre2001}. As they indicate a variation of satisfaction and effort, the corresponding associated functions (Equations~\eqref{eq:set_sat} and~\eqref{eq:set_eff}, respectively) should be adjusted (see Equations~\eqref{eq:adjust} and ~\eqref{eq:objective_fun}), when applied to a subset of requirements $\mathbf{U} \subseteq \mathbf{R}$ of requirements, as follows
\begin{equation}\label{eq:set_sat_adjusted}
    \text{sat}(\mathbf{U}) = \text{sat}(\mathbf{U}) + \text{Adjust} (\mathbf{U}, \mathbf{Y_{s}}),
\end{equation}
\begin{equation}\label{eq:set_eff_adjusted}
    \text{eff}(\mathbf{U}) = \text{eff}(\mathbf{U}) + \text{Adjust} (\mathbf{U}, \mathbf{Y_{e}}).
\end{equation}

Finally, \emph{exclusion} dependencies $\mathbf{Ex}$ indicate which requirements cannot be developed in the same product. They therefore lead to different products and consequently define different requirements selection problems that have to be dealt with independently, with their own treatment of attributes, interactions and modifications, as defined before~\cite{sagrado2015}.

\subsection{Description of a concise case study}
\label{subsec:didactic_ex}

Next, we will give an example that brings together all the above issues.  It is a project in which two clients help the engineers to identify some requirements to be included in the next release of a software project. Both clients are of equal importance to the project. Consider the situation where, in the release planning process of a specific agile project, the developers start with a set of eighth requirements $\{r_1, r_2, r_3, r_4, r_5, r_6, r_7, r_8\}$, for which the clients have estimated their revenues or benefits that define their satisfaction. The corresponding development efforts are also available.
However, the team knows that some requirements could be more risky than others; in fact, a risk estimate is available for all of them, with $r_2$ and $r_8$ being the most risky. 

Although the company needs to consider risks in this next version because the company wants to defer product features that may present some risk, this could change for future releases Perhaps, once the current release is completed and delivered, the new next release strategy, defined by the company, would need to leave out risks and take into account another requirement property, such as dissatisfaction, in order to apply the Kano model, as this approach also takes into account features that dissatisfy clients~\cite{babok2015}. We need a general framework that can be customized according to the project status and the company's strategy for the new release, while maintaining the key aspects of requirement engineering.

\begin{figure}[ht]
\footnotesize
\begin{center}

\includegraphics[width=0.4\textwidth]{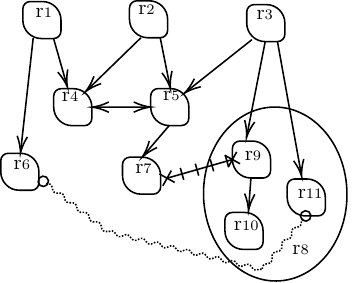}

\caption{Ideographic representation of interactions.}\label{fig:interactions}

\end{center}
\end{figure}

At the end of the elicitation sessions, not only our initial set of requirements contains eleven requirements ($q=11$) because $r_8$ has been groomed, breaking it down into three  candidate features to be treated separately in the project, $\mathbf{Re} = \{r_1, r_2, r_3, r_4, r_5, r_6, r_7, r_8, r_9, r_{10}, r_{11}\}$,  but also interactions between requirements have been available. 
The refinement interactions are represented by $\mathbf H=\{ (r_8, r_9), (r_8, r_{10}), (r_8, r_{11}) \}$. In addition, structural interactions have been identified as, $\mathbf I=\{ (r_1, r_4), (r_1, r_6), (r_2, r_4), (r_2, r_5), (r_5, r_7), (r_3, r_5), (r_3, r_8), $ $(r_3, r_9), (r_3, r_{11}), (r_9, r_{10})\}$, $\mathbf J=\{ (r_4, r_5) \}$, and
$\mathbf {Ex}=\{ (r_4, r_7) \}$. Finally, the developers estimate that $r_6$ and $r_{11}$ have a strong connection, knowing that if they are both developed at the same time, a certain amount of resources will be saved, about 10\% each. So a value-based interaction can be defined between the two requirements, 
 $\mathbf{Y_{effort}}= \{(\{r_6, r_{11}\}, 0.9) \}$. An ideograph to represent these interactions appears in the Figure \ref{fig:interactions}.

The elicitation process allows the team to define the properties to be recorded about the requirements, because on the one hand the stakeholders facilitate the quantitative values they estimate about the benefits and penalties they assign to each requirement, and on the other hand the developers have been assigned a risk for each requirement to be considered. In this way, the set of attributes available for $\mathbf{Re}$ is $\mathbf{At}=\{value, penalty, cost, risk \}$. The table \ref{tab:case} contains the values defined for this didactic example. In addition, $r_2$ collects the specification of a feature that is mandatory for the next release of the product. At this point, the team needs to set the effort bound according to the available resources and the team velocity.  For the case study, the effort bound is set at 18 effort units.

\begin{table}[htpb]
\caption{Concise case: available attributes.}\label{tab:case}

\centering
\begin{tabular}{|lrrrrrrrrrrr|}
\hline
\textbf{Requirements}    & $r_1$ & $r_2$ & $r_3$ & $r_4$ & $r_5$ & $r_6$ & $r_7$ & $r_8$ & $r_9$ & $r_{10}$&$r_{11}$\\ \hline
\textit{value client1} &  1&	1&	2&	0&	5&	0&	2&	4&	2&	1&	1     \\
\textit{value client2}   &5	&1&	1&	2&	0&	1&	1&	4&	2&	0&	2   \\
\textit{penalty client1} &0	&1	&1	&2	&2	&2	&0	&4	&0	&2	&2    \\
\textit{penalty client2} & 1	&2	&3	&2	&5	&1	&2	&7	&3	&3	&1    \\ 
\textit{satisfaction}    &6	&2	&3	&2	&5	&1	&3	&9	&5	&1	&3     \\
\textit{dissatisfaction}  & 1	&3	&4	&4	&7	&3	&2	&15	&7 &	5&	3     \\
\textit{effort}          & 3	&4	&2	&1	&4	&3	&2	&10	&3 &	2&	5     \\
\textit{risk}            & 2	&3	&1	&2	&2	&1	&2	&3	&3 &	1&	1     \\ \hline
\end{tabular}

\end{table}

\section{The \ensuremath{\mathsf{NRP}} optimization problem} 
 \label{sec:staged-approach}

Due to the computational complexity of the requirements selection problem, it has been formulated as an optimization problem with the aim of moving it from a human-based search to a machine-based search, using a variety of optimization algorithms. The typical goal is to find the ideal set of requirements that balances stakeholder satisfaction within a set of fixed effort constraints.

 The formulation of an optimization problem involves: choosing one or more optimization variables, determining the objective functions, and eliciting problem constraints~\cite{bhatti2012}. Then, a multi-objective optimization problem can be defined as minimizing (or maximizing) simultaneously multiple objective functions. More formally, 

 \begin{equation} \label{eq:genop}
    \begin{array}{ll}
         \text{minimize} & F(\mathbf{x}) = (f_1(\mathbf{x}), f_2(\mathbf{x}), \cdots, f_z(\mathbf{x}))  \\
         \text{subject to} & \mathbf{x} \in \mathbf{X}, \\
                           & f_i(\mathbf{x}) \geq l_i, \text{ for certain functions } f_i(\mathbf{x}) \in F(\mathbf{x}),\ \\
                           & f_j(\mathbf{x}) \leq g_j, \text{ for certain functions } f_j(\mathbf{x}) \in F(\mathbf{x}),\\
    \end{array}
\end{equation}

\noindent where  $F(\mathbf{x}) = (f_1(\mathbf{x}), \cdots , f_{z}(\mathbf{x}))$ is the set of multiple objective functions, $\mathbf{X}$ is the feasible set of decision vectors, $\mathbf{x} \in \mathbf{X}$ is a feasible solution, and $f_i(\mathbf{x}) \geq l_i$, $f_j(\mathbf{x}) \leq g_j$ are constraints that any $\mathbf{x}$ have to fulfill for some given constant values $l_i$ and $g_j$~\cite{coello2007}.

There is usually no feasible solution that minimizes all objective functions simultaneously. Therefore, attention is paid to feasible solutions that cannot be improved in any of the objectives without degrading at least one of the other objectives. This is the concept of \emph{dominance}. Formally, let $\mathbf{x}_1, \mathbf{x}_2 \in \mathbf{X}$ be two feasible solutions. We say that $\mathbf{x}_1$ \emph{dominates} $\mathbf{x}_2$ if the following conditions are met

\begin{equation} \label{eq:dominance}
    \begin{array}{ll}
    \text{\emph{i)} } & \forall i \in \lbrace 1, 2, \cdots, z \rbrace, \text{ } f_i(\mathbf{x}_1) \leq f_i(\mathbf{x}_2), \text{ and}\\ 
    \text{\emph{ii)} } & \exists i \in \lbrace 1, 2, \cdots, z \rbrace, \text{ } f_i(\mathbf{x}_1) < f_i(\mathbf{x}_2).
    \end{array}
\end{equation}

Thus, a feasible solution $\mathbf{\hat{x}} \in \mathbf{X}$ is \emph{Pareto optimal} if there is no other feasible solution that dominates it. The set $\mathbf{\hat{X}}$ of Pareto optimal solutions, is known as the \emph{Pareto front}.

According to this definition, an \ensuremath{\mathsf{NRP}} can be formulated as a multi-objective optimization problem, where a feasible solution is a set of requirements, and objectives and constraints are functions that relate to the attributes and interactions of the requirements that make up the feasible solution. Objectives and constraints may be linear or non-linear, and continuous or discrete in nature, but anyway they are based on the available information on requirements.  The typical formulations in the literature are:

\begin{itemize}
\item Single objective formulation~\cite{bagnall2001, van2005, sagrado2011, greer2004, sagrado2010}
\begin{equation}\label{mono}
  \begin{array}{ll}
         \text{maximize} & \text{sat}(\mathbf{U})\\
         \text{subject to} & \mathbf{U} \subseteq \mathbf{R}, \\
         & \text{eff}(\mathbf{U}) \leq B\\
         & \mathbf{U} \text{ fulfills } \mathbf{I}
         \\
    \end{array}
\end{equation}
\item  Bi-objective formulation~\cite{zhang2007, sagrado2015, dominguez2019, durillo2011, finkelstein2009}
\begin{equation}\label{multi}
  \begin{array}{ll}
         \text{maximize} & \text{sat}(\mathbf{U})\\
         \text{minimize} & \text{eff}(\mathbf{U})\\
         \text{subject to} & \mathbf{U} \subseteq \mathbf{R}, \\                
         & \text{eff}(\mathbf{U}) \leq B\\
         & \mathbf{U} \text{ fulfills } \mathbf{I}
    \end{array}
\end{equation}
\end{itemize}

However, these formulations have drawbacks. One is that they ignore real situations that arise in projects, such as the need to capture risk and dissatisfaction, or the presence of mandatory requirements, or complex interactions such as value interactions. Another is that they do not easily support changes in the target functions, as in the example (see section ~\ref{subsec:didactic_ex}) where, between iterations of the requirements selection task, it becomes necessary to focus first on risk and then on dissatisfaction.

\section{\ensuremath{\mathsf{NRP}} extended definition: \ensuremath{\mathsf{extNRP}}} 
\label{sec:formal-definition}

Due to the variety of data associated with requirements, the interactions that may exist between requirements, and the need for adaptation that arises in development projects, it is necessary to have an open generic formulation that can be instantiated throughout the evolution in a particular software project or in software engineering methodologies. Therefore, the extended version, \ensuremath{\mathsf{extNRP}}, will have to handle a non-closed set of optimisation objectives and all the attributes and constraints described in Section~\ref{sec:data}.

At this point, the set of the current agreed set of attributes associated with requirements \cite{SWEBOK2014}, $\mathbf{At}$, has been identified: \emph{satisfaction}, \emph{dissatisfaction}, \emph{effort}, \emph{risk}, \emph{price}, \emph{urgency} and \emph{instability}.
From these, functions have been defined that allow the value of attributes of individual requirements to be generalized to sets of requirements:
satisfaction (\emph{sat}, eq. \eqref{eq:set_sat}), dissatisfaction (\emph{dis}, eq.\eqref{eq:set_dis}),  effort (\emph{eff}, eq. \eqref{eq:set_eff}), risk (\emph{risk}, eq. \eqref{eq:set_risk}), price (\emph{pri}, eq. \eqref{eq:set_pri}), time sensitivity (\emph{tim}, eq. \eqref{eq:set_urg}) and instability (\emph{ins}, eq. \eqref{eq:set_ins} ). Table \ref{tab:fitness} collects a description of these functions and details of their definitions. These are the functions used in the different versions of \ensuremath{\mathsf{NRP}}.
The values of these functions can be bounded by constants which are set from the value of the attributes of all the requirements involved in the selection problem. In addition, also requirement interactions define constraints on the problem, as feasible subsets of requirements have to fulfill them.

The problem of requirements selection for the next release falls within this general framework. Formally, \ensuremath{\mathsf{extNRP}} problem consists of selecting subsets of requirements such that
\begin{equation} \label{eq:optimi}
    \begin{array}{ll}
         \text{maximize} & F^1(\mathbf{U}) = (f_1^1(\mathbf{U}), f_2^1(\mathbf{U}), \cdots, f_{z'}^1(\mathbf{U})) \\
         \text{minimize} & F^2(\mathbf{U}) = (f_1^2(\mathbf{U}), f_2^2(\mathbf{U}), \cdots, f_{z''}^2(\mathbf{U}))  \\
         \text{subject to} & \mathbf{U} \subseteq \mathbf{R}, \\
                           & f_i^1(\mathbf{U}) \geq b_i^1, \text{ for certain functions } f_i^1(\mathbf{U}) \in F^1(\mathbf{U}),\\
                           & f_j^2(\mathbf{U}) \leq b_j^2, \text{ for certain functions }f_j^2(\mathbf{U}) \in F^2(\mathbf{U}), \\
                           & \mathbf{U} \text{ fulfills } \mathbf{H}, \mathbf{I}, \mathbf{J}, \neg \mathbf{Ex},
    \end{array}
\end{equation}
where the set of objective functions to maximize $F^1(\mathbf{U})$ can be any subset in $\{$\emph{sat}, \emph{tim}$\}$, 
the set of objective functions to minimize $F^2(\mathbf{U})$ any subset in $\{$\emph{dis}, \emph{eff}, \emph{pri}, \emph{risk}, \emph{ins}$\}$.  Note that these sets of functions can be tailored to suit the company, the project or the specific release. 
The constant values $b_i^1$ and $b_j^2$ for the objective functions set constraints for the objective functions and are computed from the value of the attributes of all the requirements involved in the selection problem 
and any feasible subset of requirements have to fulfill refinement, implication, combination interactions and no exclusion. Value interactions have been considered previously, by adjusting the value of the attributes of the requirements, and that of the objective functions.

Now, we are able to answer the first research question,  \emph{RQ1}. The requirement selection problem defined by \ensuremath{\mathsf{extNRP}} is a multi-objective optimization problem in which the objective functions are based on the attributes (i.e., $a_i \in \mathbf{At}$) of the requirements and on value interactions. Then, for each selected attribute in \ensuremath{\mathsf{extNRP}}, an objective function $f_{a_i}$  
is defined according to Equation~\eqref{eq:objective_fun}. 
The problem constraints are defined by interactions and boundary values for the attributes. It should be noticed that, in this way, the  \ensuremath{\mathsf{extNRP}} formulation (see eq. \eqref{eq:optimi}) is adaptable and open to the incorporation of new objective functions and constraints in those projects that according to the methodology and tools applied use different requirements attributes.

 Returning to our concise case study in Section~\ref{subsec:didactic_ex}, applying  the \ensuremath{\mathsf{extNRP}} definition gives us the following optimization problem:

\begin{equation}\label{exa}
   \begin{array}{ll}
         \text{maximize} & \text{sat}(\mathbf{U})\\
         \text{minimize} & \text{dis}(\mathbf{U}), \text{eff}(\mathbf{U}), \text{risk}(\mathbf{U})\\
         \text{subject to} & \mathbf{U} \subseteq \mathbf{R}, \\ 
         & \text{eff}(\mathbf{U}) \leq 18,\\
         & \mathbf{U} \text{ fulfills } \mathbf{H}, \mathbf{I}, \mathbf{J}, \neg \mathbf{Ex}.
    \end{array}
\end{equation}

\noindent 
If we compare the \ensuremath{\mathsf{extNRP}} (eq. \ref{exa}) and \ensuremath{\mathsf{NRP}} (eq. \ref{multi}) definitions for this problem, we can see that the \ensuremath{\mathsf{extNRP}} includes all requirements data, all objective functions and all interactions between requirements that are available in the problem, while the \ensuremath{\mathsf{NRP}} definition does not. Table ~\ref{tab:compara} shows this comparison at requirement data level.

\vspace{1ex}
\noindent\fbox{%
  \parbox{\textwidth}{
 This new unified formulation for \ensuremath{\mathsf{NRP}} is the answer to \emph{RQ1} because it allows the inclusion of no limited number of optimization objectives and constraints involved in the decision about which requirements, within a set of elicited requirements, are going to be covered by the new release that is going to be developed. Previous multi-objective formulations of the \ensuremath{\mathsf{NRP}} focus on a fixed number of attributes (i.e., effort or and satisfaction) and constraints, without taking into account what is the information about requirements that is really available in a software development project, thus limiting their scope.
  }%
}

\begin{table}[htpb]

\caption{Concise case: comparison of \ensuremath{\mathsf{NRP}} and \ensuremath{\mathsf{extNRP}} definitions at the level of requirements data.}

\label{tab:compara}     
\centering
\begin{tabular}{|p{35ex}|p{4ex}|r|c|c| }
\hline
 \textbf{Requirements} & \textbf{$R^j$} & & \ensuremath{\mathsf{NRP}} & \ensuremath{\mathsf{extNRP}}\\
\hline
$r_1, r_3,$ $r_4, r_5,$ $r_6$, $r_7,$ $ r_9,$ $ r_{10},r_{11}$ && & x& x\\
$r_2$ &&Mandatory & & x\\
$r_8$ & $R^{8}$ & $\{r_{9}, r_{10}, r_{11} \}$   & & x\\   
\hline
\multicolumn{3}{|p{55ex}|}{$\mathbf I=\{ (r_1, r_4), (r_1, r_6), $ $(r_2, r_4),$ $ (r_2, r_5),$ $ (r_5, r_7),$ $ (r_3, r_5), $ $(r_3, r_8), $ $(r_3, r_9), $ $(r_3, r_{11}),$ $ (r_9, r_{10})\}$
 }  & x& x\\
\hline
\multicolumn{3}{|p{50ex}|}{
$\mathbf J=\{ (r_4, r_5) \}$}   &  & x\\
 \hline
\multicolumn{3}{|p{50ex}|}{
$\mathbf {Ex}=\{ (r_7, r_9) \}$}  & & x\\
 \hline
\multicolumn{3}{|p{50ex}|}{
$\mathbf{Y_{effort}}= \{(\{r_6, r_{11}\}, e_6*0.9$ \& $e_{11} *0.9) \}$
}   & & x\\
\hline
\end{tabular}
\end{table}

\begin{sidewaystable}
\centering
\caption{Objective functions used in \ensuremath{\mathsf{extNRP}}. }
\label{tab:fitness}

\begin{minipage}{\textheight}
\begin{tabular}{|p{17ex}|p{30ex}|c|p{45ex}|}
\hline
Name & Description & Definition & Function components meaning \\
\hline
Satisfaction & Stakeholders' advantage as a result of a requirement implementation & $ \text{sat}(\mathbf{ U}) = \displaystyle\sum_{r_i \in \mathbf{U }} s_i$ &  $s_j = \sum_{i=1}^{m}w_i * v_{ij}$, $v_{ij}$ represents benefit that stakeholder $c_i$ defines for $r_j$ \\
Effort & Estimated person-hours needed to develop a requirement & $\text{eff}(\mathbf{ U}) = \displaystyle\sum_{r_i \in \mathbf{U }} e_i$ & When effort is broken in components (development tasks) $e_i= \sum_{j} ec_{ij}$, $j$ represents the different tasks  \\
Risk & An estimate of an uncertain event that, if it were to occur, the requirement could lose its potential benefit & $\text{risk} (\mathbf{U}) = \displaystyle\sum_{r_j \in \mathbf{U}} rk_j$ &  $rk_{j}$ represents the risk estimated for the requirements $r_j$ \\
Time sensitivity & Expiration date for the requirements &$\text{tim}(\mathbf{ U}) = \displaystyle\sum_{r_j \in \mathbf{U }} n_j$ &   $  n_j = \sum_{i=1}^{m}w_i* u_{ij}$, $u_{ij}$ represents the urgency that stakeholder $c_i$ estimates for $r_j$ \\
Price & Development cost &$  \text{pri}(\mathbf{ U}) =\displaystyle \sum_{r_j \in \mathbf{U }} pr_j$ &$pr_j$ represents the development cost in monetary units \\
Instability & It represents how unstable a requirement is &$\text{ins}(\mathbf{ U}) = \displaystyle\sum_{r_j \in \mathbf{U }} is_j$ & $is_j = \sum_{i=1}^{m}w_i * vl_{ij}$, $vl_{ij}$ represents the volatility that stakeholder or developer $i$ defines for $r_j$ \\
\hline
\end{tabular}
\end{minipage}

\end{sidewaystable}

\section{\ensuremath{\mathsf{extNRP}} usage} 
\label{sec:analysis}

When faced with the selection of requirements in a given software development project, we are confronted with a challenging problem due to the complexity and the many different relationships between the elements involved, such as the high amount of stakeholders, the variety and heterogeneous nature of the variables to be reviewed or the uncertainty of the data used to solve it~\cite{Ruhe2010}. These are the reason why 
the \ensuremath{\mathsf{extNRP}} formulation (see eq. \eqref{eq:optimi}) has to be tailored to the specific characteristics of the project.

The first thing to do is to determine the set of objective functions (i.e., $F^1(\mathbf{U})$ and $F^2(\mathbf{U})$) and constraints (i.e., the requirements interactions and the constant values $b_i^1$, $b_j^2$ that bound the objective functions) from the available data on the attributes of the requirements involved in the project. These will be the elements to be considered in the definition (see eq. \eqref{eq:optimi}) of the \ensuremath{\mathsf{extNRP}} problem to be solved. For instance, 
once the set of mandatory requirements involved in the selection problem, has been identified $\mathbf{R} \subseteq \mathbf{Re}$ after removing, the set $\mathbf{P}$ of regulatory requirements and considering changes due to interactions, the stakeholders' satisfaction associated with an individual requirement can be obtained by aggregating the satisfaction values that each stakeholder assigns to it~\cite{bagnall2001,zhang2007}.

Finally, among these alternatives (i.e., Pareto optimal solutions) the development team, even in discussion with some stakeholders,  will have to choose which one to implement. Human experts in charge of making decisions need to be supported by additional analysis on the optimization results because of the black box nature of the optimization algorithms~\cite{du2014}. Although there are very well-known quality indicators to measure the performance of algorithms~\cite{nuh2021} and of the Pareto fronts (such as hypervolume or spread~\cite{ali2020}), these metrics quantify aspects that are out of the software project scope. To make this choice, they need other qualitative or quantitative factors that are closer to the world of the project, such as the current market policies of the company where the software is being developed.  Even the mathematical theories that base their models on preference and utility functions consider it quite difficult to incorporate the preferences of stakeholders and teams when selecting a single solution point after the problem has been solved~\cite{arora2012}. 

Unfortunately, there is no roadmap for deciding which solution is the most appropriate for a given a project situation. Some authors propose to explain the results using quality indicators closer to the problem domain~\cite{aguila2016}, machine learning algorithms from the calculated decision attributes (metrics or indicators) in the Pareto front~\cite{du2014}, or binary search trees~\cite{HUJAINAH2021}.

\subsection{Analysis of \ensuremath{\mathsf{extNRP}} solutions}
\label{sec:quality_indicators}
In fact the analysis of the solutions obtained could be considered as a separate problem not always easy to solve, which could be treated using complex techniques such as a Pareto-optimal solution ranking and selection of the best solution in multi- and many-objective optimization problems~\cite{rao2021} or ranking based on knee solutions~\cite{choa2020}.
 
For \ensuremath{\mathsf{extNRP}} we propose the use  of an a posteriori strategy \cite{ferreira2007} supported by   quality indicators that, correctly displayed by some kind of tool, guide decision-makers when comparing solutions~\cite{aguila2016}. Besides the data resulting from the optimization algorithms (such as the number of requirements in a solution, the detailed list of requirements, and the values achieved by the solutions in the objective functions) some other helpful indicators can be calculated. 

We define three types of indicators: \emph{ratio indicators, coverage indicators} and \emph{calculated indicators}.  The first group, \emph{ratio indicators}, represents relationships between values of the objective functions. Let $\mathbf{U} \subseteq \mathbf{R}$ the solutions under analysis, then
\begin{itemize}
\item \emph{Productivity}, 
\begin{equation}\label{eq:productivity}
   \text{prod}(\mathbf{U})=\text{sat}(\mathbf{U})/\text{eff}(\mathbf{U}),
\end{equation}
is the benefit obtained by the solution expressed in terms of how much satisfaction is obtained per unit of effort.
 
\item \emph{Dirtiness}, 
\begin{equation}
   \text{dirt}(\mathbf{U}) = \text{dis}(\mathbf{U}) / \text{sat}(\mathbf{U}),
\end{equation}
 is the degree of stakeholder dissatisfaction per unit of benefit achieved. 
\end{itemize}

\noindent Not all the pairs of attributes give decision-makers a useful clue when they triage the Pareto front, because a specific pair could have no meaning for the project.

The second group, \emph{coverage indicators}, deals with multivalued attributes. These indicators represent how the aggregated value in the solution covers the individual values originally assigned from the different sources. 
For instance, the measure of the amount covered by a solution with respect to everything raised by the stakeholder (i.e., stakeholder fairness), is called \emph{coverage}~\cite{aguila2016}. Thus,
given a stakeholder $c_i \in \mathbf{C}$, this measure associated to a solution $\mathbf{ U} \subseteq \mathbf{R}$ with respect to all the requirements valuated by her/him is
\begin{equation}
    \text{stcov}_i(\mathbf{U})=  {\sum_{j \in {\mathbf{U}}} v_{ij}}/ {\sum_{j \in {\mathbf{R}}}v_{ij}},
\end{equation}
where $v_{ij}$ is the value that the stakeholder $c_i$ assigns to  requirement $r_j$.
 
Effort can also be a multivalued attribute. In this case the coverage for each specific development task (i.e., effort component $t$) can be calculated as
\begin{equation}
 \text{efcov}_t (\mathbf{U})= {\sum_{r_j \in \mathbf{U}} ec_{tj}} /  {\sum_{r_j \in \mathbf{R}} ec_{tj}}.   
\end{equation}
\noindent where $ec_{tj}$ represents the effort value for requirement $r_j $ in category $t$ defined according to the project methodology.

Finally, \emph{calculated indicators}, which are closer to the project management domain, are computed using relevant project data and solution-specific measures. A representative of this group is the \emph{Squandering} indicator, which measures the percentage of wasted resources for a given solution
\begin{equation}
     \text{squa}(\mathbf{U})={( B - \text{effort}(\mathbf{U}) }) / {B}.
\end{equation}

The set of quality indicators can be extended according to the attributes described in Section~\ref{sec:attributes} without affecting \ensuremath{\mathsf{extNRP}} definition or usage. However, the use of too many indicators can hinder rather than help decision-making. The general rule should be the simpler and more complete the analysis, the better. For example, in situations where too many stakeholders are involved, the use of coverage (i.e. \emph{stcov}) is not recommended. But, pairwise comparisons of solutions focusing on specific data or the use of graphical representations (such as those of interactions~\cite{aguila2016}) should be encouraged, without exhaustively exploring indicators. 
This approach to the solution analysis task would not only help in selecting the set of requirements to be included in the software product, but could also support the negotiation processes with stakeholders to clarify the release goals.

In summary, the \ensuremath{\mathsf{extNRP}} application process comprises three distinct phases. In the first, based on the available data on the attributes of the requirements involved in the project, the \ensuremath{\mathsf{extNRP}} optimization problem is determined by identifying the objective functions and constraints. Next, an optimization algorithm is used to obtain the set of Pareto optimal solutions of \ensuremath{\mathsf{extNRP}}, subsets of requirements that could be incorporated into the next software release. Finally, after an analysis of these alternatives, the one to be implemented is chosen. Each phase is independent enough to be adapted and customized according to the specific project and the data available for the requirements, which facilitates the usability of \ensuremath{\mathsf{extNRP}}.\\

\vspace{1ex}
\noindent\fbox{%
  \parbox{\textwidth}{
The \ensuremath{\mathsf{extNRP}} usage process is the answer to the second research question, \emph{RQ2}: each phase could be managed and customized independently according to the particular project and taking into account the available attributes for the requirements. In this way, \ensuremath{\mathsf{extNRP}} can be adapted to any development project and facilitates the identification of the set of requirements to be included into the next software release.

}}

\subsection{Analysis of the concise case study}

Let us return to the example in section~\ref{subsec:didactic_ex}, with the set of requirements $\mathbf{Re} = \{r_1, r_2, r_3, r_4, r_5, r_6, r_7, r_8, r_9, r_{10}, r_{11}\}$, where $r_2$ is mandatory and $r_8$ have been groomed, and the interactions (see Fig. \ref{fig:interactions}) expressed by the sets $\mathbf H=\{ $$(r_8, r_9), $$(r_8, r_{10}), (r_8, r_{11}) \}$, $\mathbf I=\{ (r_1, r_4), (r_1, r_6), $ $(r_2, r_4),$ $ (r_2, r_5),$ $ (r_5, r_7),$ $ (r_3, r_5), $ $(r_3, r_8), $ $(r_3, r_9), $ $(r_3, r_{11}),$ $ (r_9, r_{11})\}$, $\mathbf J=\{ (r_4, r_5) \}$,
$\mathbf {Ex}=\{ (r_7, r_9) \}$ and   $\mathbf{Y_{effort}}= \{(\{r_6, r_{11}\}, 0.9) \}$. 

It is worth noting that the development team prefers a groomed backlog (i.e. by taking into account refinement interactions in $\mathbf{H}$) and this sets the level at which requirements are considered, thus defining $\mathbf{R} = \{r_1, r_3, r_4, r_5, r_6, r_7, r_9, r_{10}, r_{11}\}$, where $r_2$ is not included because as it is mandatory it will always has to be included in the next release.  Due to combination interactions $\mathbf{J}$, we need to define a new requirement $r_{4+5}$, which replaces in $\mathbf{R}$ the combined requirements, so $\mathbf{R} = \{r_1, r_3, r_{4+5}, r_6, r_7, r_9, r_{10}, r_{11}\}$. These changes in the set of requirements have to be reflected in the interactions. In our case, they affect the implication interactions as follows: $\mathbf I=\{ (r_1, r_{4+5}), (r_1, r_6),$ $ (r_{4+5}, r_7), (r_3,r_{4+5}), $ $ (r_3, r_9), (r_3, r_{11}),  (r_9, r_{10})\}$. Also, each exclusion in $\mathbf{Ex}$ forces us to consider two different set of requirements, $\mathbf{R'} = \{r_1, r_3, r_{4+5}, r_6, r_7, r_{11} \}$ or 
$\mathbf{R''} = \{r_1, r_3, r_{4+5}, r_6, r_9, r_{10}, r_{11} \}$, to start from when defining the next release. Value interactions $\mathbf{Y_{effort}}$ affect the value of effort and, in this case the necessary adjustments are given by $\text{eff}({{r_i}}) = \text{eff}(r_i)*0.9$, where $r_i \in \{r_6, r_{11}\}$. Finally, the effort bound decreases, $B'=B-\text{eff} (r_2)=14$, because the requirement $r_2$ is mandatory. The adjusted values for the requirements data in the groomed backlog are collected in Table~\ref{tab:casead}.

\begin{table}[htpb]
\caption{Concise case: requirements and attributes for the groomed backlog.}
\label{tab:casead}

\centering
\begin{tabular}{|lrrrrrrrr|} 
\hline
\textbf{Requirements}     & $r_1$  & $r_3$ & $r_{4+5}$ & $r_{6}$ & $r_7$  & $r_9$ & $r_{10}$ & $r_{11}$ \\ \hline
\textit{satisfaction}     & 6&	3&	7&	1&	3&	5&	1&	3         \\
\textit{dissatisfaction}  & 1	&4	&11&	3&	2&	7&	5&	3       \\
\textit{effort}           & 3	&2	&5	&3	&2	&3	&2	&5     \\
\textit{risk}             &2	&1	&2	&1	&2	&3	&1	&1     \\ \hline

\end{tabular}

\end{table}

With the adjusted requirements data for the groomed backlog, applying the \ensuremath{\mathsf{extNRP}} definition gives us the following optimization problem:

\begin{equation}\label{exa_groomed_backlog}
   \begin{array}{ll}
         \text{maximize} & \text{sat}(\mathbf{U})\\
         \text{minimize} & \text{dis}(\mathbf{U}), \text{eff}(\mathbf{U}), \text{risk}(\mathbf{U})\\
         \text{subject to} & \mathbf{U} \subseteq \mathbf{R} = \{r_1, r_3, r_{4+5}, r_6, r_9, r_{10}, r_{11} \}, \\ 
         & \text{eff}(\mathbf{U}) \leq 14,\\
         & \mathbf{U} \text{ fulfills } \mathbf{H}, \mathbf{I}, \mathbf{J}, \neg \mathbf{Ex}.
    \end{array}
\end{equation}

\noindent Next, we use an optimization algorithm (in our study case we use an exhaustive search due to the small number of requirements involved) to obtain the set of 17 Pareto optimal solutions for this \ensuremath{\mathsf{extNRP}}. What remains is for decision-makers to analyze the solutions and select the one to be implemented in the next release. And they usually prefer to analyze solutions close to the resource limit $B=14$. Table~\ref{tab:qualitycase} lists the five solutions in the Pareto front that are closest to this limit, together with the values (the best values are marked with an asterisk) that each solution achieves in the quality indicators used to compare them.

\begin{table}[htpb]

\caption{Concise case: handpicked solutions and quality indicators to compare them.}
\label{tab:qualitycase}
\centering
\begin{tabular}{|cccccccccc| }
\hline

 Id& eff & sat & dissat & risk&  N.req. & product. & dirtiness& ${stcov_1}$ & ${stcov_2}$ \\
 \hline
$sol_{1}$ & 12*         &19\text{ } &18\text{ } &7\text{ }  &4\text{ } &1.58\text{ }  &0.95\text{ }	&71\%*	      &64\%\text{ }\\
$sol_{2}$ & 13\text{ }	&13\text{ }	&11*     	&5*	        &4\text{ } &1.00\text{ }  &0.85*	    &29\%\text{ } &64\%\text{ }\\
$sol_{3}$ & 13\text{ } 	&16\text{ } &20\text{ } &8\text{ }	&5*	       &1.23\text{ }  &1.25\text{ }	&43\%\text{ } &64\%\text{ } \\
$sol_{4}$ & 13\text{ }	&17\text{ } &19\text{ } &6\text{ }	&4\text{ } &1.31\text{ }  &1.12\text{ } &57\%\text{ }  &64\%\text{ } \\
$sol_{5}$ & 13\text{ }  &21*        &23\text{ }	&8\text{ }	&4\text{ } &1.62*	      &1.10\text{ } &71\%*\text{ }&71\%* \\
\hline
\end{tabular}

\end{table}

Solution $sol_5$ has the best satisfaction and achieves a better rating in 3 of the 4 quality indicators, shows better productivity and better stakeholder coverage, but has the highest dissatisfaction and risk score of the five solutions analyzed. On the other hand, solution $sol_2$ has the lowest dissatisfaction, risk and dirtiness, but is not better in terms of satisfaction, effort or stakeholder coverage than $sol_5$. All these factors make the decision difficult and need to be taken into account when negotiating the release goal.

\section{Coverage of previous \ensuremath{\mathsf{NRP}} instances}
\label{sec:instances}

\begin{figure}[ht]
\footnotesize
\begin{center}

\includegraphics[width=\textwidth]{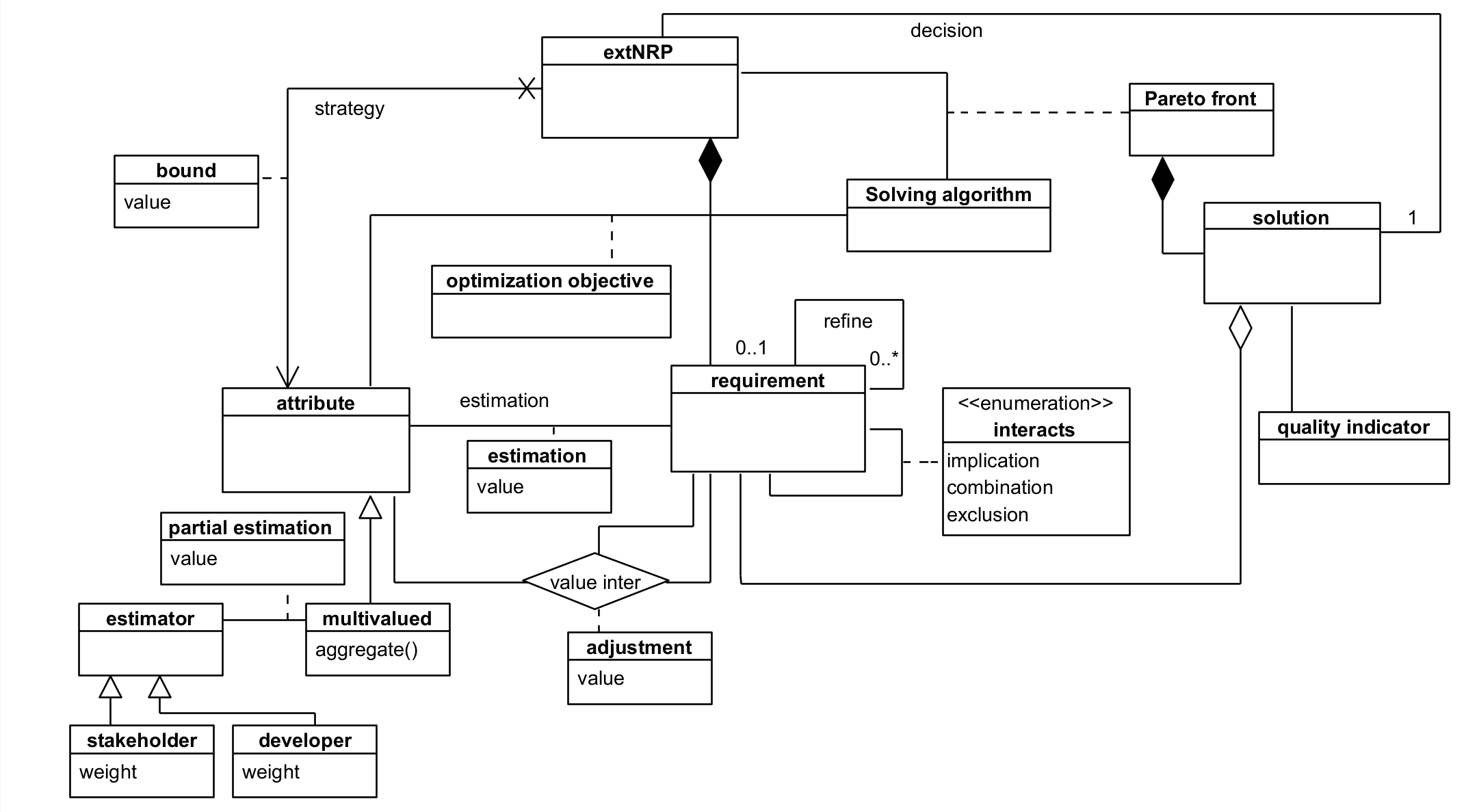}
\caption{\ensuremath{\mathsf{extNRP}} Metamodel.}\label{fig:metamodel}
\end{center}
\end{figure}

After having defined the \ensuremath{\mathsf{extNRP}} (see Section~\ref{sec:formal-definition}) and its usage (see Section~\ref{sec:analysis}), it remains to be studied how previous \ensuremath{\mathsf{NRP}} instances can be handled with it. In order to achieve this, we will apply the \ensuremath{\mathsf{extNRP}} usage process by adapting it to each specific \ensuremath{\mathsf{NRP}} instance depending on the data associated with the requirements that are available. Table~\ref{tab:fitness} lists the objective functions that appear in the various projects discussed below,  and Table~\ref{tab:instance} provides a summary of the characteristics of each instance at the requirements data level (see Section~\ref{sec:data}).    Therefore, the process we will follow for a given \ensuremath{\mathsf{NRP}} instance and its associated set of requirements data consists in: defining the \ensuremath{\mathsf{extNRP}} optimization problem, applying an optimization algorithm to obtain the set of Pareto optimal solutions for the \ensuremath{\mathsf{extNRP}}, and perform a basic analysis of the solutions in order to choose the one to be implemented in the next release.  Figure \ref{fig:metamodel} shows the metamodel that describes our extended formulation, including the concepts and issues that can be managed for \ensuremath{\mathsf{extNRP}}.  
We are not primarily concerned about the set of solutions returned by optimization algorithms when applied to \ensuremath{\mathsf{extNRP}}, but on \ensuremath{\mathsf{extNRP}} usage. That is the reason why the choice of the optimization algorithm in each \ensuremath{\mathsf{NRP}} instance is done based on available data on requirements. We want to get a set of solutions in a short time and with enough quality to perform an analysis, without regarding that other optimization algorithm could obtain a better solutions set. A comparison of optimization algorithms applied to \ensuremath{\mathsf{extNRP}} is out of the scope of this work.

The basic solution analysis process applied starts by calculating, for each solution, some quality indicators from among those proposed in Section \ref{sec:quality_indicators}. Each solution is then scored with one point for each indicator and objective function for which it achieves the best value. The choice of the solution to be implemented in the next software release will be made among those solutions that have obtained the maximum score.

\begin{table} [!t]

\caption{Characteristics of previous \ensuremath{\mathsf{NRP}} instances: number of stakeholders (\#stk) and requirements (\#req), presence of mandatory requirements (mand.), requirement interactions ($\mathbf H (levels)$, $\mathbf I$, $\mathbf J$, $\mathbf {Ex}$,   $\mathbf {Y_a}$), number of constraints (\#B) and objective functions (\#obj).}
\label{tab:instance}
\centering
\begin{tabular}{|lcccccccccc| }
\hline
 Instance
 & \#stk &\#req&   mand. & $\mathbf H (levels)$  &$\mathbf I$  & $\mathbf J$  & $\mathbf {Ex}$  &  $\mathbf {Y_a}$  &   \#\textit{B} &\#obj   \\
 \hline
Motorola &4& 40   &  $\checkmark$ &       -  &       - &       - &      - &      - &      1 & 2 \\ 
Classic  &5& 20   &      -      &  -  & $\checkmark$  & $\checkmark$ &    -   & -        & 1 &2\\
MSLite  &-&  16   &      -      &             &        -           &-& -   & -           & 1&3\\
RALIC   &76& 10   &     -       & $\checkmark$ (3) &       - &      - &      - & - &         1 &2\\ 
        &76& 48   &     -       & $\checkmark$ (3) &       - &      - &      - & - &         1 &2\\ 
        &76& 104  &     -       & $\checkmark$ (3)&       - &      - &      - & - &         1 &2\\ 
WordProc&4 & 8    & -           & $\checkmark$ (2)& $\checkmark$& $\checkmark$ &-&   $\checkmark$ & 1&3\\
        &4 & 50   & -           & $\checkmark$ (2)& $\checkmark$& $\checkmark$ &-&   $\checkmark$ & 1&3\\
Theme RP &9& 5     &  -          & $\checkmark$ (2)& $\checkmark$& $\checkmark$ &-&   - & 2&6\\
        &9& 25    &  -          & $\checkmark$ (2)& $\checkmark$& $\checkmark$ &-&   - & 2&6\\
\hline
\end{tabular}

\end{table}

\subsection{Motorola project}

baThis instance comprises a set of 40 requirements ($\mathbf {Re}$) which are candidates to be included in the possible future evolution of the software controlling a mobile telecommunications device.  Four experts discussed the components and decided that five entries belong to the regulatory requirements category (see Section~\ref{sec:attributes}) so they should be mandatory. That is, the set  $\mathbf R$ contains the remaining 35 components, and the attribute values were estimated by the four experts defining the classical Motorola data set, widely applied in \ensuremath{\mathsf{NRP}} research works~\cite{baker2006}.

The attributes on which data are collected are satisfaction and effort. Thus, a feasible solution $\mathbf{U} \subseteq \mathbf{R}$, should maximize stakeholders' satisfaction and minimize development effort as these are the objective functions involved (see Table~\ref{tab:fitness}). There are no interactions between the requirements, but a given development effort bound $B$ has to be preserved, giving rise to a constraint $\text{eff}(\mathbf{U}) \leq B'$.  The reason for this is that five requirements are mandatory and their effort need to be taken into account. Therefore, the \ensuremath{\mathsf{extNRP}} definition for this problem is 
\begin{equation}\label{35nrp}
  \begin{array}{ll}
         \text{maximize} & \text{sat}(\mathbf{U})\\
         \text{minimize} & \text{eff}(\mathbf{U})\\
         \text{subject to} & \mathbf{U} \subseteq \mathbf{R}, \\                              & \text{eff}(\mathbf{U}) \leq B'
    \end{array}
\end{equation}
\noindent This problem formulation using \ensuremath{\mathsf{extNRP}} covers the bi-objective \ensuremath{\mathsf{NRP}} formulation proposed in \cite{zhang2007}.

Before applying the optimization algorithm, which in this case is a greedy algorithm, the values assigned by the four stakeholders to each of the 35 requirements have to be aggregated into a single value to compute the requirement satisfaction, $s_i$. Besides, the effort bound has been set to 21\% of the total effort ($B'=1415$). As a result of the execution of the greedy algorithm we obtain a set of 86 Pareto optimal solutions. Of these, the ones that are really useful are those that are close to the resource limit, $B'$. So for the analysis, only a  handpicked set of solutions closer to $B'$ should be managed. The choice of the number of solutions to be studied relies in the human experts and can be changed at any time. Table~\ref{tab:qualitymoto} shows the selected solution sets, the objective functions values, the quality indicators used and, as satisfaction is a multivalued attribute, the coverage of each stakeholder. Best values are marked with an asterisk.

\begin{table} [!t]
\caption{Handpicked solutions and quality indicators for Motorola project}
\label{tab:qualitymoto}
\centering
\begin{tabular}{|cccccccccc| }
\hline
 Id\footnotemark[1]& sat &  eff. & N.req. & Prod. &  Squ. &  $stcov_{1}$ & $stcov_{2}$ & $stcov_{3}$ & $stcov_{4}$ \\
 \hline
$sol_{1310}$ & 49*        & 1310 & 24*         & 0.036*        & 7.0\%\text{ } & 60\%* & 83\%\text{ } & 15\%\text{ } & 100\%*\\
$sol_{1390}$ & 49*        & 1390 & 24*         & 0.035\text{ } & 1.4\%\text{ } & 60\%*       & 83\%\text{ } & 25\%*& \text{ 89\% }\\
$sol_{1400}$ & 47\text{ } & 1400 & 21\text{ }  & 0.033\text{ } & 0.7\%\text{ } & 50\%\text{ } & 83\%\text{ } & 15\%\text{ } & 100\%*\\
$sol_{1410}$ & 48\text{ } & 1410 & 24*         & 0.034\text{ } & 0.0\%*        & 60\%* & 83\%\text{ } & 25\%*& \text{ 84\% }\\
\hline
\end{tabular}

\footnotetext[1]{for identifier $sol_{i}$, i represents the effort of the solution}
* best value in the indicator/column
\end{table}

For this project, the solution $sol_{1310}$ achieves the best value for five quality indicators,  so it seems to be the most appropriate. However, if according to decision makers it is recommended to spend all of the project resources, $sol_{1390}$ or $sol_{1410}$ could be good alternatives  because they achieve maximum effort without wasting resources. It is precisely in these situations where pairwise comparisons are useful~\cite{aguila2016}.

\subsection{Classic \ensuremath{\mathsf{NRP}} project with interactions}

The classic \ensuremath{\mathsf{NRP}} project ~\cite{greer2004} focuses on satisfaction and effort.
It comprises 20 requirements and 5 costumers with different importance. The data collected on this project includes customers' weights ($w_i$),  values assigned by each customer to requirements ($v_{ij}$) and the effort associated to each requirement ($e_j$). Also, implication ($\mathbf{I}$) and combination ($\mathbf{J}$) interactions between requirements are considered:
$\mathbf I=\{(r_4,r_8)$, $(r_4,r_{17})$
 $(r_8,r_{17})$,($r_9,r_3)$,$(r_9,r_6)$,
 $(r_9,r_{12})$,
$(r_9,r_{19})$, $(r_{11},r_{19})\}$,  
$\mathbf J=\{(r_3,r_{12})$,$(r_{11}, r_{13})\}$.

From these data, we first have to obtain the satisfaction $s_j$ of each requirement $r_j$ by applying equation \eqref{eq:req_sat}. Interactions also produce changes in the values of the attributes associated with the requirements. Thus, the combination interactions cause the initial set of requirements $\mathbf{Re}$ to be reduced from 20 to 18, by grouping $(r_3, r_{12})$ into a single requirement $r_{3+12}$, and $(r_{11}, r_{13})$ into $r_{11+13}$, defining the set $\mathbf{R}$ of requirements in the selection problem. Consequently, the satisfaction and effort values of the resulting requirements after the combination will be obtained by adding those of the original requirements. It is also necessary to replace in the implication interactions set the occurrences of the original requirements by those resulting after the combination~\cite{sagrado2015}. That is, $r_3$ and $r_{12}$ are replaced by $r_{3+12}$, and $r_{11}$ and $r_{13}$ by $r_{11+13}$, resulting in a new set of implication interactions $\mathbf{I'}$. The resource bound is set to 25 effort units ($B=25$), the 30\% of the total effort.

Thus, a feasible solution $\mathbf{U} \subseteq \mathbf{R}$, should maximize stakeholders' satisfaction and minimize development effort as these are the objective functions involved(see Table~\ref{tab:fitness}). There are implications interactions between the requirements, and a development effort bound $B = 25$ has to be preserved. Therefore, the \ensuremath{\mathsf{extNRP}} definition for this problem is 
\begin{equation}\label{20enrp}
   \begin{array}{ll}
         \text{maximize} & \text{sat}(\mathbf{U})\\
         \text{minimize} & \text{eff}(\mathbf{U})\\
         \text{subject to} & \mathbf{U} \subseteq \mathbf{R}, \\                              & \text{eff}(\mathbf{U}) \leq B,\\
         & \mathbf{U} \text{ fulfills } \mathbf{I'}
    \end{array}
\end{equation}
\noindent This problem formulation using \ensuremath{\mathsf{extNRP}} incorporates interactions and extends the bi-objective \ensuremath{\mathsf{NRP}} formulation proposed in \cite{zhang2007}.

In this case, we apply an exhaustive search algorithm \cite{sagrado2015} to obtain the set of Pareto optimal solutions. And as in the previous project, for the analysis, only a handpicked set of 4 solutions closest to the resource limit, $B$ are considered. Table~\ref{tab:quality20} shows the selected solution sets, the objective functions values (i.e., satisfaction and effort), the quality indicators used (i.e., number of requirements, productivity and squandering) and, as satisfaction is a multivalued attribute, the coverage of each stakeholder. Best values are marked with an asterisk.

\begin{table}[!t]
\caption{Handpicked solutions and quality indicators for the Classic \ensuremath{\mathsf{NRP}} project}
\label{tab:quality20}
\centering
\begin{tabular}{|cccccc| }
\hline

 Id\footnotemark[1]& sat &  eff & N.Req & Prod. &  Squa. \\
 \hline
$sol_{22}$ & 466\text{ } & 22\text{ } & \text{ 9 } & 21.18*        & \text{12\% } \\
$sol_{23}$ & 458\text{ } & 23\text{ } & \text{ 9 } & 19.91\text{ } & \text{ 8\% } \\
$sol_{24}$ & 497*        & 24\text{ } & \text{ 9 } & 20.71\text{ } & \text{ 4\% } \\
$sol_{25}$ & 482\text{ } & 25*        & 10*        & 19.28\text{ } & \text{ 0\%}* \\
\hline
\multicolumn{6}{|c|}{Stakeholder coverage}\\
\hline
 Id\footnotemark[1]&  ${stcov_1}$ & ${stcov_2}$ & ${stcov_3}$ & ${stcov_4}$& ${stcov_5}$\\
 \hline
$sol_{22}$ & 46.7\%\text{ } & 52.4\%\text{ } & 56.2\%\text{ } & 47.7\%\text{ } & 54.5\%\text{ } \\
$sol_{23}$ & 43.5\%\text{ } & 54.1\%\text{ } & 54.7\%\text{ } & 47.7\%\text{ } & 51.5\%\text{ } \\
$sol_{24}$ & 46.8\%*        & 57.4\%*        & 57.8\%*        & 52.3\%*        & 57.6\%*\\
$sol_{25}$ & 46.8\%*        & 55.7\%\text{ } & 57.8\%*        & 50.8\%\text{ } & 54.5\%\text{ } \\
\hline
\end{tabular}

\footnotetext[1]{for identifier $sol_{i}$, i represents the effort of the solution}
* best value in the indicator/column
\end{table}

The solution with effort 24, $sol_{24}$, scores 6 points. It has the best satisfaction and  stakeholders coverage. It is followed by the solution with effort 25, $sol_{25}$, with 5 points. This is the one that includes more requirements, and it does not squander any resources, but it has less satisfaction and stakeholders coverage than $sol_{24}$. However, although the $sol_{22}$ has less satisfaction than the two previous ones, it is the one that makes the best use of its resources, achieving the best productivity value. From this analysis, $sol_{24}$ is the first option to be selected and implemented in the next software release. However, if other considerations arise, such as a historical decrease in the team velocity, it could be dangerous for the next iteration to choose a solution with a smaller resource gap (squandering), since if the trend is confirmed the release could fail. Decision makers can opt in favor of $sol_{22}$ to keep 12\% of the resources reserved, reaching acceptable values of satisfaction.

\subsection{MSLite System project}

This project includes user stories and acceptance test cases for the MSLite~\cite{sangwan2020} software system. It is a system that automatically monitors and controls the internal functions of a building, such as heating, ventilation, air conditioning, access, and safety. It comprises a set $\mathbf{R}$ with 16 requirements, for each of which satisfaction, dissatisfaction and effort values are provided. Also the following set of implication interactions is defined: $\mathbf I=\{ (r_1,r_2), $ $(r_1,r_{12}),$ $ (r_1,r_{14}),$ $(r_1,r_{15}),$ $(r_1,r_{16}),$ $(r_2,r_{6}),$ $ (r_2,r_{15}),$ $(r_3,r_{4}),$ $(r_3,r_{5}),$ $  (r_3,r_{7}),$ $(r_4,r_{13}),$ $(r_4,r_{14}),$ $ (r_6,r_{7}),$ $(r_8,r_{1}),$ $(r_8,r_{2}),$ $ (r_8,r_{3}),$ $(r_8,r_{4})$, $(r_8,r_{5}),$ $(r_8,r_{6}),$ $(r_8,r_{7}),$ $(r_8,r_{15}),\}$.
Because all attributes (i.e., satisfaction, dissatisfaction and effort) are directly estimated for each requirement, and no combination interactions are described, no initial computation for attribute values is needed. 

Thus, a feasible solution $\mathbf{U} \subseteq \mathbf{R}$, should maximize stakeholders' satisfaction and, minimize dissatisfaction and development effort, as these are the objective functions involved (see Table~\ref{tab:fitness}). There are implications interactions between the requirements, and a given development effort bound $B$, which has been set at 21 effort units, has to be preserved. Therefore, the \ensuremath{\mathsf{extNRP}} definition for this problem is 

\begin{equation}\label{16enrp}
   \begin{array}{ll}
         \text{maximize} & \text{sat}(\mathbf{U})\\
         \text{minimize} & \text{dis}(\mathbf{U}), \text{eff}(\mathbf{U})\\
         \text{subject to} & \mathbf{U} \subseteq \mathbf{R}, \\                              & \text{eff}(\mathbf{U}) \leq B,\\
         & \mathbf{U} \text{ fulfills } \mathbf{I}
    \end{array}
\end{equation}

In this project, the low number of requirements makes it possible to use an exhaustive search algorithm to obtain the Pareto front. In the analysis of the 6 chosen solutions closest to the resource limit, $B$, coverage indicators cannot be used, as we do not have multivalued attributes.
Instead ratio (i.e. productivity, dirtiness, annoyance)  and calculated (i.e. squandering) indicators are used. In this project analysis we have included an additional ratio indicator, \emph{annoyance}, 
\begin{equation}
\text{ann}(\mathbf{U}) = \text{diss}(\mathbf{U}) / \text{eff}(\mathbf{U}),
\end{equation}
which measures how much disturbance is included for each effort unit.  Table~\ref{tab:quality16} shows the selected solution sets, their values for objective functions (i.e., satisfaction, dissatisfaction and effort) and the quality indicators. 

\begin{table}
\caption{Solutions, objective functions and quality indicators for MSLite project}
\label{tab:quality16}
\centering
\begin{tabular}{|ccccccccc| }
\hline
Id \footnotemark[1] & sat &  dis & eff & N.Req.	& dirt &	prod &	annoy.	& squa \\
\hline

$sol_{(31.57.21)}$& 31\text{ } & 57\text{ } & 21\text{ } & \text{ 8 } &	1.84\text{ } &	1.48\text{ } &	2.71\text{ } &	\text{ 0\%}* \\
$sol_{(35.56.19)}$& 35\text{ } & 56\text{ } & 19*        & \text{ 8 } &	1.60\text{ } &	1.84\text{ } &	2.95\text{ } &	10\%\text{ } \\
$sol_{(36.54.20)}$& 36\text{ } & 54*        & 20\text{ } & \text{ 8 } &	1.50\text{ } &	1.80\text{ } &	2.70*        &	\text{ 5\% } \\
$sol_{(54.74.19)}$& 54\text{ } & 74\text{ } & 19*        & 10*        &	1.37\text{ } &	2.84\text{ } &	3.89\text{ } &	10\%\text{ } \\
$sol_{(57.70.20)}$& 57*        & 70\text{ } & 20\text{ } & 10*        &	1.23*        &	2.85*        &	3.50\text{ } &	\text{ 5\% } \\
$sol_{(52.74.21)}$& 52\text{ } & 74\text{ } & 21\text{ } & 10*        &	1.42\text{ } &	2.48\text{ } &	3.52\text{ } &	\text{ 0\%}* \\
\hline
\end{tabular}

\footnotetext[1]{for identifier $sol_{s.d.e}$, s, d, and e represent respectively satisfaction, dissatisfaction, and effort of the solution}
* best value in the indicator/column
\end{table}

The most appropriate alternative should have
low dirtiness (i.e., dissatisfaction included per satisfaction unit), high productivity (i.e., satisfaction per effort unit), and low annoyance. In situations like this, quality indicators act as extra knowledge sources to the decision makers. From the set of chosen solutions, solution $sol_{(57.70.20)}$ is the one with the highest score of 4 points. It is the best alternative as it has the highest satisfaction, the lowest dirtiness and the highest productivity.

\subsection{Replacement Access, Library and ID Card project}

The aim of this software project was to improve the existing access control system at University College London (UCL). The project combined different UCL access control mechanisms into one, such as access to library and fitness
center, eliminating the need for a separate library registration process for UCL ID card holders~\cite{lim2011}.

There are 138 requirements in this project, defining the original set $\mathbf{Re}$. They are arranged in three levels, as \emph{refinement} interactions, $\mathbf{H}$, have been defined. Each of these levels, from the most general to the most specific, contains 10, 45 and 85 requirements, respectively. The effort needed to develop the requirements is provided at each of the these levels. The stakeholder group consists of 76 UCL members. Each stakeholder $c_i$ indicates which requirements $r_j$ her/him is interested in by scoring them with a value $v_{ij}$ in the interval $[-1, 5]$. Thus, in order to set the satisfaction value $s_j$ for each requirement $r_j$, we have to apply equation \eqref{eq:req_sat} and the importance of each stakeholder $c_i$ has to be quantified as a weight $w_i$. A quantification of stakeholders is necessary~\cite{babar2015} because their weight is not quantitatively defined in this project. Fortunately, influence among stakeholders is defined by a social network based on a recommendation system, in which the influence, that according to her/his personal opinion, the stakeholder $c_j$ gives to another stakeholder $c_i$ in the RALIC project, is expressed by a value \emph{influence}$(c_i, c_j)$ in the interval [1, 8]. From these influence values, we can calculate the weight of any of the $m$ (i.e., 76) stakeholders. The weight $w_i$ of the stakeholder $c_i$ is obtained as
\begin{equation}\label{ralic}
    w_i=\mbox{ $\sum_{j=1}^{m} \text{\emph{influence}}(c_i,c_j)$.}
\end{equation}
\noindent Note that other methods can be used to obtain the weights of stakeholders~\cite{lim2011, zanaty2021}, but each leads to a different optimization problem as weights have a direct influence on satisfaction values.

Next, the \emph{refine} interactions $\mathbf{H}$ have to be taken into account and the set of requirements $\mathbf{R}$ to be considered in the selection problem has to be determined accordingly, in order to obtain the values of the attributes (satisfaction and effort) of the requirements. 
On this occasion, as the set of requirements $\mathbf R=\{a, b, c, d, e, f, g, h, i, j\}$ in the selection problem, the 10 that are at the highest level of the hierarchy defined by $\mathbf{H}$ have been chosen. A feasible solution $\mathbf{U} \subseteq \mathbf{R}$ in the RALIC project, should maximize stakeholders' satisfaction and minimize development effort, there are no  interactions left between the requirements, but a given development effort bound $B$ (set at 50\% of the total resources needed) has to be preserved, giving rise to the constraint $\text{eff}(\mathbf{U}) \leq B$. With all these considerations the \ensuremath{\mathsf{extNRP}} definition for this problem is the same given in equation \eqref{35nrp}.

\begin{table}[ht]
\caption{Solutions, objective functions and quality indicators for RALIC project}
\label{tab:quality10}
\centering
\begin{tabular}{|cccccc| }
\hline

Id \footnotemark[1]& sat& eff& N.req.	&	prod & squa \\
\hline
$sol_{10163}$&283701.44*&	10163*	&6\phantom{*}	&27.91*	&0.76\%\phantom{*} \\
$sol_{10184}$&102720.73\phantom{*}&	10184\phantom{*}&	4\phantom{*}	&10.09\phantom{*}&	0.55\%\phantom{*}\\
$sol_{10197}$&\phantom{0}80896.48\phantom{*}&	10197\phantom{*}&	3\phantom{*}&	\phantom{0}7.93\phantom{*}&	0.42\%\phantom{*}\\
$sol_{10206}$&242865.03\phantom{*}&	10206\phantom{*}&	7*&	23.80\phantom{*}	&0.34\%\phantom{*}\\
$sol_{10229}$&103356.16\phantom{*}	&10229\phantom{*}	&4\phantom{*}	&10.10\phantom{*}	&0.11\%\phantom{*}\\
$sol_{10239}$&244468.08\phantom{*}	&10239\phantom{*} &7*	&23.88&	0.01\%*\\
\hline
\end{tabular}

\footnotetext[1]{for identifier $sol_{i}$, i represents the effort of the solution}
* best value in the indicator/column
\end{table}

The reduced number of requirements allows the use of an exhaustive search algorithm to obtain the Pareto front. Table~\ref{tab:quality10} shows the set of solutions close to the effort limit that has been analyzed. Due to the high number of stakeholders, the use of coverage indicators is ruled out. In addition to the objective functions, only productivity and squandering indicators are used in the analysis. The choice would be $sol_{10163}$ as it scores the highest with 3 points, with the best values for the objective functions (satisfaction and effort), the highest productivity and a controlled squandering.

\subsection{Word processing software project}

This case study~\cite{agarwal2014} corresponds to a project planned in $2008$ that has been derived from the commonly known word processing tool MS Word. It comprises $58$ requirements, arranged at two levels, and $81$ functional interactions that have been elicited from $4$ weighted stakeholders with weights $\mathbf W =\{9, 3, 5, 7\}$. Table~\ref{tab:50requirementdepe} shows the requirements and the interactions that have been elicited for this problem. Value interactions have been expressed as a percentage of increase instead of an absolute value.

\begin{table}  [!t]
\caption{Word processing project: requirements and interactions}
\label{tab:50requirementdepe}     
\centering
\begin{tabular}{|p{8ex}|l|l| }
\hline
 \textbf{Reqs} & \textbf{$r^j$}  \\
\hline
File & $r^{file}$ & $\{r_{f1}, r_{f2}, r_{f3}, r_{f4}, r_{f5}, r_{f6}, r_{f7}, r_{f8}, r_{f9}, r_{f{10}}, r_{f{11}}  ,r_{f{12}} \}$\\
Edit & $r^{edit}$& $\{r_{e1}, r_{e2}, r_{e3}, r_{e4}, r_{e5}, r_{e6}, r_{e7}, r_{e8}, r_{e9}, r_{e{10}} \}$ \\
View&$r^{view}$& $\{r_{v1}, r_{v2}, r_{v3}, r_{v4}, r_{v5} \}$ \\
Insert&$r^{insert}$&  $\{r_{i1}, r_{i2}, r_{i3}, r_{i4}, r_{i5} \}$ \\
Format&  $r^{format}$& $\{r_{m1}, r_{m2}, r_{m3}, r_{m4}, r_{m5}, r_{m6} \}$ \\
Tools& $r^{tools}$ &$\{r_{t1}, r_{t2}, r_{t3}, r_{t4}, r_{t}, r_{t6} \}$ \\
Data& $r^{data}$ & $\{r_{d1}, r_{d2}, r_{d3}, r_{d4}, r_{d5}, r_{d6} \}$ \\
Help& $r^{help}$ & $\{r_{h1}, r_{h2} \}$ \\
\hline
\multicolumn{3}{|c|}{ \textbf{I, implication functional interactions}}\\
\hline
\multicolumn{3}{|p{90ex}|}{ $\mathbf I=\{(r_{f1},r_{f4}),$
 $(r_{f1},$ $r_{f5}),$  $(r_{f1},$ $r_{f6}),$ $(r_{f1}$ $,r_{f8}), $ $(r_{f1},$ $r_{f9}),$ $(r_{f1},$ $r_{f{10}}),$ $(r_{f1},$ $r_{f{11}}), $ $(r_{f1},$ $r_{f{12}}),$ $(r_{f1},$ $r_{v2}),$ $ (r_{f1},$ $r_{v3}),$ $(r_{f1},$ $r_{v4}),$ $(r_{f1},$ $r_{v5}),$ $(r_{f1},$ $r_{t1}),$  $(r_{f1},$ $r_{t2}), (r_{f2},$ $r_{f3}), $ $(r_{f2},$ $r_{f6}),$ $ (r_{f2},$ $r_{f7}),$  $ (r_{f2},$ $r_{f9}), $ $(r_{f2},$ $r_{f{10}}),$ $ (r_{f2},$ $r_{f{11}}), $ $(r_{f2},$ $r_{e1}), $  $(r_{f2},$ $r_{e2}), $ $(r_{f2},$ $r_{v2}), $ $(r_{f2},$ $r_{v3}),$ $ (r_{f2},$ $r_{v4}), $ $(r_{f2},$ $r_{v5}),$ $ (r_{f2},$ $r_{i1}),$ $ (r_{f2},$ $r_{i2}), $ $ (r_{f2},$ $r_{i3}),$ $ (r_{f3},$ $r_{f{12}}),$ $ (r_{f4},$ $r_{f{12}}),$   $  (r_{f5},$ $r_{f{12}}), $ $(r_{f6},$ $r_{f{12}}), (r_{f7},$ $r_{f{10}}), $ $ (r_{f7},$ $r_{{12}}),  (r_{f8},$ $r_{f{9}}), $ $(r_{f8},$ $r_{f{12}}), $ $(r_{f9},$ $r_{f{12}}),$ $ (r_{f{10}},$ $r_{f{12}}),$  $  (r_{f{11}v},$ $r_{F{12}}), (r_{e7},$ $r_{e5}), $ $ (r_{e8},$ $r_{e9}),$  $  (r_{v2},$ $r_{f{9}}), (r_{v5},$ $r_{v2}), $ $ (r_{v5},$ $r_{v3}), (r_{i1},$ $r_{i4}), $ $ (r_{i1},$ $r_{i5}),$  $ (r_{i2},$ $r_{i4}), $ $ (r_{i2},$ $r_{i5}),$ $ (r_{m1},$ $r_{m2}), $ $ (r_{m1},$ $r_{m3}), $ $ (r_{m1},$ $r_{m4}), $ $ (r_{m1},$ $r_{m5}),$ $ (r_{m2},$ $r_{m5}),$ $ (r_{m3},$ $r_{m5}), (r_{m4},$ $r_{m5}),$  $ (r_{t1},$ $r_{t2}),$ $ (r_{t2},$ $r_{t3}),$ $ (r_{t6},$ $r_{t1}), $ $ (r_{t6},$ $r_{t2}), $ $ (r_{t6},$ $r_{t3}), $  $ (r_{t6},$ $r_{t4}), $ $ (r_{t6},$ $r_{t5})$ $
 \}$
 }\\
\hline
\multicolumn{3}{|c|}{ \textbf{J, combination functional interactions}}\\
\hline
\multicolumn{3}{|p{90ex}|}{
$\mathbf J=\{ (r_{f2},r_{f4}),$ $ (r_{f4},r_{f5}), $ $(r_{e3},r_{e4}),$ $ (r_{e3},r_{e5}),$ $ (r_{e4},r_{e5}), $ $(r_{v5},r_{i1}),$ $ (r_{d1},r_{d2}), $ $(r_{d1},r_{d3}), $ $(r_{d2},r_{d3}),$ $ (r_{h1},r_{h2})\}$}\\
 \hline
 \multicolumn{3}{|c|}{ $\mathbf Y_{e}$, \textbf{value interactions for effort}}\\
\hline
\multicolumn{3}{|p{90ex}|}{
$ {Y_{e}}=\{ ({r_{e3}, r_{e4}, r_{e5}, r_{e6}, r_{e7}, r_{e{10}}}, +e_i*0.4), ({r_{e1}, r_{e3}, r_{e4}, r_{e5}, }$ ${r_{e8}, r_{e}}, +e_i*0.4),$ $ ({r_{v1}, r_{v2}, r_{v3}},  +e_i*0.3),$ $ ({r_{i3},r_{i5}},  +e_i*0.2),$ $ ({r_{t1}, }$ ${ r_{t2}, r_{t3}}, r_{h1}+r_{h2}*0.3),$ $ ({r_{t4}, r_{t5}}, +e_i*0.25), $ $({r_{d1}, r_{d4}, r_{d5}}, +e_i*0.3) \}$
}\\
\hline
\end{tabular}

\end{table}

The list of available attributes for requirements comprises: value, urgency, and effort. Whereas the effort needed to develop each requirement has been estimated in person-hours by the development team, each stakeholder $c_i$ gives her/his estimation on the value ($v_{ij}$) and urgency ($u_{ij}$) of each requirement $r_j$ and the values of these multivalued attributes has to be calculated using equations \eqref{eq:req_sat} and \eqref{eq:req_urg}, for \emph{satisfaction} and \emph{time sensitivity} respectively. Estimates of requirement attribute values provided by stakeholders and developers have been made at the lower level of granularity of the set $\mathbf{Re}$ of original requirements. Consequently, the set $\mathbf{R}$ of requirements to be considered in the selection problem is reduced to the 50 requirements at the lowest level of $\mathbf{Re}$. 

As there are combination interactions $\mathbf{J}$, they have to be considered, reducing the number of requirements in $\mathbf{R}$ to 42 by combining those ones that are involved in a combination interaction~\cite{sagrado2015}. For the combined requirements, the values of effort, satisfaction and time sensitivity have to be adjusted by adding those values of the requirements involved in the combination interactions. Also, the set $\mathbf{I}$ of implication interactions has to be rearranged in a new set $\mathbf{I'}$ to take into account the combination of requirements.

In this project, a feasible solution $\mathbf{U} \subseteq \mathbf{R}$, should maximize stakeholders' satisfaction and time sensitivity and minimize development effort as these are the objective functions involved (see Table~\ref{tab:fitness}). But, as there are value interactions $\mathbf{Y_e}$ defined for effort (see Table~\ref{tab:50requirementdepe}), the effort objective function has to be adjusted following equation \eqref{eq:objective_fun}, so the effort for a feasible solution $\mathbf{U} \subseteq \mathbf{R}$ becomes
$$
\text{eff}(\mathbf{U}) = \text{eff}(\mathbf{U}) +  \text{Adjust}(U, Y_e).
$$
There are also implications interactions $\mathbf{I'}$ between the requirements, and a development effort bound $B$ (which has been set to 709, 50\% of the effort of all requirements in $\mathbf{R}$) has to be preserved. Therefore, the \ensuremath{\mathsf{extNRP}} definition for this problem is 

\begin{equation}\label{50enrp}
   \begin{array}{ll}
         \text{maximize} & \text{sat}(\mathbf{U}), \text{tim}(\mathbf{U})\\
         \text{minimize} & \text{eff}(\mathbf{U})\\
         \text{subject to} & \mathbf{U} \subseteq \mathbf{R}, \\                              
         & \text{eff}(\mathbf{U}) \leq B,\\
         & \mathbf{U} \text{ fulfills } \mathbf{I'}
    \end{array}
\end{equation}

For this \ensuremath{\mathsf{extNRP}} a hill climbing algorithm has been used to obtain a Pareto front with 133 solutions. Table~\ref{tab:quality} shows the set of 5 solutions closest to the limit $B$ that have been analyzed, the objective functions and quality indicators used. Solution $sol_{(2500.2480.708)}$ gets the highest score of 9 points and can be a good candidate for the next release. However, in a negotiation with stakeholders, $sol_{(2340,2338,699)}$ and  $sol_{(425,2391,695)}$  present the same coverage as $sol_{(2500,2480,708)}$ and could be alternatives to consider. Should problems arise in the project, the lack of squandering in $sol_{(2500,2480,708)}$ could be a disadvantage.

\begin{table} [ht]
\caption{Quality indicators for  results in word processing project}
\label{tab:quality}
\centering
\begin{tabular}{ |cccccccc| }
\hline
 Id  \footnotemark[1] & sat & tim & eff & N.Req & $prod_{sat}$ & $prod_{tim}$ & squa \\
 \hline
 $sol_{(2182.2189.683)}$ & 2182\text{ }	&2189\text{ }&	683* &	14\text{ }	&3.195\text{ }	&3.205\text{ }&	4\%\text{ }  \\
 $sol_{(2342.2332.692)}$ & 2342\text{ }	& 2331\text{ }	&692	\text{ }&15\text{ } &3.384\text{ }	&3.368\text{ }	& 2\%\text{ } \\
$sol_{(2425.2391.695 )}$ &  2425\text{ }&	2391\text{ }	&695\text{ }&15\text{ }&	3.489\text{ }	&3.440\text{ }&	2\%\text{ } \\ 
$sol_{(2340.2338.699)}$ & 2340\text{ }	&2338\text{ }&	699\text{ }	&15\text{ }&	3.348\text{ }&	3.345\text{ }&	1\%\text{ } \\ 
$sol_{(2500.2480.708)}$ & 2500* &	2480*	&708\text{ }	&16*&	3.531* &3.503*	&0\%* \\
\hline
\end{tabular}

\begin{tabular}{ |ccccc| }
 &  $stcov_1$ & $stcov_2$ & $stcov_3$ & $stcov_4$\\
 \hline
 $sol_{(2182.2189.683)}$ &  11\%	\text{ } & 10\%	\text{ }& 10\% \text{ }& 10\% \text{ }\\
$sol_{(2342.2332.692)}$  &  11\% \text{ } & 10\% \text{ }& 10\% \text{ }& 10\% \text{ }\\
$sol_{(2425.2391.695)}$ & 	13\% * & 10\% \text{ }& 12\%*& 13\%*\\
$sol_{(2340.2338.699)}$   &  13\% * &	10\% \text{ }& 12\%*& 13\%*\\
$sol_{(2500.2480.708)}$  &  13\% * &	10\%\text{ } & 12\%*& 13\%*\\
\hline

\end{tabular}

\footnotetext[1]{for identifier $sol_{s.t.e}$, s, d, and e represent respectively satisfaction, time sensitivity, and effort of the solution}
* best value in the indicator/column
\end{table}

\subsection{Theme-based Release Planning project}

The requirements in these project~\cite{karim2014} are arranged in two levels. The first one includes 5 requirements that are broken down into 25 low-level requirements. Associated to the low-level requirements there are only \emph{implication} and \emph{combination} interactions. Table~\ref{tab:25complex} shows in detail requirements and interactions defined in this project. 

\begin{table}  [!t]
\caption{Theme-based Release Planning project: Requirements Interactions}
\label{tab:25complex}      
\centering
\begin{tabular}{ |l|l|l| }
\hline
 \textbf{Reqs} & \textbf{$r^j$}&  \\
\hline
Feature modeling & $r^{feat}$ & $\{r_{a1}, r_{a2}, r_{a3}, r_{a4}, r_{a5} \}$\\
Planning & $r^{planning}$& $\{r_{b1}, r_{b2}, r_{b3}, r_{b4}, r_{b5} \}$ \\
Analysis of priorities&$r^{anal}$,& $\{r_{c1}, r_{c2}, r_{c3}, r_{c4} \}$ \\
Import and export&$r^{imp-exp}$&  $\{r_{d1}, r_{d2}, r_{d3}, r_{d4}, r_{d5} \}$ \\
Analysis of plans&  $r^{ana-plan}$& $\{r_{e1}, r_{e2}, r_{e3}, r_{e4}, r_{e5}, r_{e6} \}$ \\
\hline
\multicolumn{3}{|c|}{ \textbf{I, Implication functional interactions}}\\
\hline
\multicolumn{3}{|p{90ex}|}{ $\mathbf I=\{(r_{c1},r_{c3}),$
 $(r_{d2},r_{d1}),$  $(r_{d2},r_{d3}),$ $(r_{d2},r_{d4}), $ $(r_{d2},r_{d5}),$ $(r_{e1},r_{d5}),$
 $(r_{e2},r_{d5}),$ $(r_{e3},r_{d5}),$   $(r_{e4},r_{d5}),$  $(r_{e5},r_{d5}),$  $(r_{e6},r_{d5}) \}$
 }\\
\hline
\multicolumn{3}{|c|}{ \textbf{J, Combination functional interactions}}\\
\hline
\multicolumn{3}{|p{62ex}|}{
$\mathbf J=\{ (r_{a4},r_{a5}),$ $ (r_{a3},r_{a4}), $ $(r_{a3},r_{a5}),$ $ (r_{b2},r_{e1}) \}$}\\
 \hline
\end{tabular}
\end{table}

The set $\mathbf{At}$ of requirements attributes includes \emph{satisfaction}, \emph{dissatisfaction}, \emph{effort}, \emph{price}, \emph{instability} and \emph{prevalence}. For each low-level requirement $r_j$, the value of the multivalued attributes \emph{satisfaction, dissatisfaction} and \emph{instability} are calculated from the estimates that each customer $c_i$ provides for each of them by applying equations  \eqref{eq:req_sat}, \eqref{eq:req_dis} and \eqref{eq:req_ins}, respectively. Regarding the multivalued attribute effort,
six components have been considered: C++ Backend developer, Java frontend development, Testing, Project management, Quality assurance and Requirements analysis. For each low-level requirement $r_j$, the development team gives an estimation $ec_{ij}$ for each effort component and then the requirement effort $e_j$ is computed using equation \eqref{eq:req_eff_categories}. In addition, as an attribute also associated to the resources, the \emph{cost/price} ($pr_j$) of development of each requirement is provided, expressed in thousands of dollars. As a novelty, this project includes a new attribute, the \emph{frequency of use}, $pv_j$, of a requirement $r_j$. It is defined based on the estimations provided by stakeholders as
\begin{equation}
 pv_j = \sum_{i=1}^{m}w_i * fq_{ij},
 \end{equation}
\noindent where $fq_{ij}$ represents the frequency of use that stakeholder $c_i$ assigns to $r_j$. From the frequency of use values we can compute the \emph{prevalence} of a set of requirements $\mathbf{U} \subseteq \mathbf{R}$ as
\begin{equation}
\text{pre}(\mathbf{U }) = \displaystyle\sum_{r_j \in \mathbf{U }} pv_j
\end{equation}

The set $\mathbf{R}$ of requirements in the selection problem is obtained after taking into account the combination interactions in $\mathbf{J}$, which reduce the low-level set of requirements to 23: requirements $\{r_{a3}, r_{a4}, r_{a5}\}$ conform a new requirement $r_{a3+a4+a5}$ and requirements $\{r_{b2}, r_{e1}\}$ define the new $r_{b2+e1}$. The attribute values for these new requirements are obtained by adding those of the combined requirements. Also, the set $\mathbf{I}$ of implication interactions has to be rearranged accordingly in a new set $\mathbf{I'}$ to take into account these changes in the requirement set.

Thus, a feasible solution $\mathbf{U} \subseteq \mathbf{R}$, should maximize stakeholders' satisfaction and prevalence, and minimize dissatisfaction, price, instability and development effort, as these are the objective functions involved (see Table~\ref{tab:fitness}). There are implications interactions between the requirements, and development bounds for price $B_{\text{pri}}$ and effort $B_{\text{eff}}$ (which have been set at 4800 for price and 8604 for effort) have to be preserved. Therefore, the \ensuremath{\mathsf{extNRP}} definition for this problem is 
\begin{equation}\label{eq:25complex}
   \begin{array}{ll}
         \text{maximize} & \text{sat}(\mathbf{U}), \text{pre}(\mathbf{U})\\
         \text{minimize} & \text{dis}(\mathbf{U}), \text{pri}(\mathbf{U}), \text{ins}(\mathbf{U}), \text{eff}(\mathbf{U}) \\
         \text{subject to} & \mathbf{U} \subseteq \mathbf{R}, \\                               & \text{pri}(\mathbf{U}) \leq B_{\text{pri}},   \\
                           & \text{eff}(\mathbf{U}) \leq B_{\text{eff}}, \\
                           & \text{eff}(\mathbf{U}) \leq B_{\text{eff}},
    \end{array}
\end{equation}

The Pareto front with 192 solutions for this problem has been obtained using a greedy algorithm. Table~\ref{tab:quality22} shows the set of 5 solutions closest to the effort limit $B_{\text{eff}}$ that have been analyzed, the objective functions and quality indicators used. The latter include \emph{efficacy} that represents units of satisfaction per unit of development cost and \emph{wastefulness} that represents the no used cost, because they are significant indicators for decision making in this project. The solution $sol_{2544}$ gets the highest score of 6 points and can be a good candidate for the next release as it reaches the highest satisfaction and prevalence, with the highest productivity and the lowest squandering and waste. An alternative, with lower price, is $sol_{2431}$ which has the lowest dissatisfaction, instability, annoyance and dirtiness.

\begin{table}
\caption{Quality indicators Theme-based planning project}
\label{tab:quality22}
\centering
\footnotesize
\begin{tabular}{ |cccccccc| }
\hline
Id  \footnotemark[1] &sat&	dis&	ins&	pre	&pri&	 eff &N.req\\
\hline
 $sol_{2544}$ & 2544*	& 1889\text{ }		& 2689\text{ }		& 1586*	& 3150\text{ }		& 7000\text{ }		& 8*\\
 $sol_{2445}$ & 2445\text{ }		 & 1702\text{ }	& 2507\text{ }		& 1581\text{ }		& 2500\text{ }		& 6305\text{ }		& 7\text{ }	\\
 $sol_{2431}$ & 2431\text{ }		& 1681*	& 2439*	& 1481\text{ }		& 2800\text{ }		& 6495\text{ }		& 7\text{ }	\\
 $sol_{2378}$ & 2378\text{ }		& 1697\text{ }		& 2486\text{ }		& 1585\text{ }		& 2400\text{ }		& 6375\text{ }	& 7\text{ }	\\
 $sol_{2369}$ & 2369\text{ }	 & 1771\text{ }	& 2454\text{ }		& 1375\text{ }		& 2350*	& 6345*	& 7\text{ }	\\ 
  \hline
  Id  \footnotemark[1] &Product&	Squandering	&Anoyance&	Dirtiness&	Waste &	Efficiency & \\
\hline
 $sol_{2544}$& 1.305*	 &  10\%*  &	 0.875\text{ }		& 0.743\text{ }		&34\% *	&0.808\text{ }	& \\
 $sol_{2445}$& 1.297\text{ }	 &	13\%\text{ }	  &  0.788\text{ }		& 0.696	\text{ }	& 48\%\text{ }	&0.978\text{ }	&\\
 $sol_{2431}$& 1.296\text{ }	 &	13\% \text{ }	 & 0.778*	&0.691*&	42\%	\text{ }	&0.868\text{ }	&\\
 $sol_{2378}$ & 1.303\text{ }	 &	16\%\text{ }		& 0.786\text{ }	&0.714\text{ }		&50\%\text{ }		&0.991\text{ }	&\\
  $sol_{2369}$& 1.247\text{ }	& 12\%\text{ }		& 0.820\text{ }		& 0.748\text{ }		& 51\%\text{ }		&1.008*&\\ 
\hline 

\end{tabular}

\footnotetext[1]{for identifier $sol_{i}$, i represents the satisfaction of the solution}
* best value in the indicator/column

\end{table}

\subsection{Discussion on  versatility of \ensuremath{\mathsf{extNRP}}}

 All the \ensuremath{\mathsf{NRP}} instances studied have different characteristics: number of requirements, attributes considered, interactions between requirements, objective functions, and constraints. Therefore, the \ensuremath{\mathsf{extNRP}} formulation has to face different optimization problems. 
 The set of constraints is also diverse: the types of requirement interactions considered in each project are different, and so are the values that bound the objective functions. All these characteristics influence the decision on which set of requirements should be selected to be developed in the next software release, which can be handled with the aid of appropriate quality indicators. 
 
 Despite all this diversity in the projects, the proposed \ensuremath{\mathsf{extNRP}} formulation has been successfully applied in all cases, even when new attributes are incorporated, the number of objective functions is high, and specific quality indicators are required in the analysis. The versatility of our method makes it a valuable asset in problem solving and decision making for the selection and prioritization of requirements, allowing a more flexible and efficient approach to a wide range of situations.

The capabilities of the new formulation in a development project are limited by the data available about the requirements and the estimates that can be made. For companies, these data and estimates are difficult to obtain. Therefore, if automatic methods and tools were available to obtain them (e.g. to establish the stability of a requirement by measuring the number of changes or commits made to its specification, or to define the interest/value of stakeholders based on the number of comments made on a requirement) for each project, it would encourage the use of \ensuremath{\mathsf{extNRP}}. This would allow companies to spend more time on decision-making tasks rather than estimation tasks.

In any case, regardless of whether a project has more or fewer requirements attributes, the potential of \ensuremath{\mathsf{extNRP}} lies in its adaptability to incorporate them and address the requirement selection problem, and in its ability to accommodate company strategies. These adaptations can be made between different versions or in a complex decision situation during the requirements selection task to perform a what-if analysis. That is, using \ensuremath{\mathsf{extNRP}} companies can test different strategies by generating and solving multiple optimization problems before making the final decision on what the set of requirements to be developed will be.

From a practical point of view, requirement engineering processes are carried out using requirement engineering tools~\cite{carrillo2012}. They allow stakeholders and developers to work together in eliciting and managing requirements. These requirement specification data should be the starting point of the requirements selection task, but classical models are difficult to adapt because the stored data does not always allow teams to define requirements satisfaction and effort. The embedding of \ensuremath{\mathsf{extNRP}} in these tools will allow the adaptation to the real data available and, the capacity to test different company policies. In addition, the \ensuremath{\mathsf{extNRP}} provides a framework based on quality indicators which, when presented graphically, could help decision-makers to compare candidate solutions and strategies.
\\

\vspace{1ex}
\noindent\fbox{%
  \parbox{\textwidth}{
The \ensuremath{\mathsf{extNRP}} formulation has demonstrated its applicability  in all the \ensuremath{\mathsf{NRP}} instances studied (\emph{RQ3})  and its ability to be adapted to different needs (i.e. new requirement attributes and relationships, objective functions and constraints) now and in the future. The usage shows that the \ensuremath{\mathsf{extNRP}} definition and the analysis of the solution phases are heavily dependent on the specific project, including the methodologies and working products managed on it.
}}

\section{Conclusions}
\label{sec:conclusions}

Deciding which, within a set of elicited and quantified requirements, are going to be considered is a strategic process in software development to which search based software engineering has paid attention. However, because of the evolution both in Software Engineering and optimization knowledge areas, the twenty year old definition of the problem should be improved.

This work proposes a general formulation, \ensuremath{\mathsf{extNRP}}, for the next release problem, which allow the use of a no limited set of optimization objectives based on the different requirements attributes elicited for the project at hand. Precisely, the \ensuremath{\mathsf{extNRP}} usage process is based on the available project data associated with the requirements. It comprises three distinct phases. First, the \ensuremath{\mathsf{extNRP}} optimization problem is defined by identifying the objective functions and constraints. It is in this initial phase that changes in the values of the requirements attributes due to interactions are taken into account and constraints are defined, either as limits to the values of the objective functions or as interactions between requirements.
Next, the set of Pareto optimal solutions for \ensuremath{\mathsf{extNRP}} is obtained by using a optimization algorithm.
Finally, a small set of solutions in the Pareto front is analyzed in order to chose the one  to be included in the next release. The development team, even in discussion with some stakeholders, will have to choose which one is the subset of requirements to implement.  For \ensuremath{\mathsf{extNRP}} we have proposed the use of three types of quality indicators (ratio, coverage and calculated indicators) that can be used as an aid to decision-makers when comparing such solutions to make the final decision.  Each phase is independent enough to be adapted and
customized according to the specific project.

The \ensuremath{\mathsf{extNRP}} formulation has proven its applicability and versatility in the \ensuremath{\mathsf{NRP}} instances studied. Each instance varies in complexity. We have addressed projects with different numbers of requirements, attributes, and interactions, covering all the alternatives that have been proposed in the literature so far. The process followed in \ensuremath{\mathsf{extNRP}} provides the necessary flexibility to cope with the selection of requirements in all situations encountered. Even this formulation has made it possible to incorporate new attributes and objective functions, which together with appropriate quality indicators (some defined for the specific case) make it easier to decide which set of requirements should be selected to be developed in the next software release.

\section*{Acknowledgments}

This research has been supported by the Data, Knowledge and Software Engineering (DKSE) research group (TIC-181) of the University of Almer\'{\i}a. Research and Transfer Plan of the University of Almeria, funded by “Consejería de Universidad, Investigación e Innovación de la Junta de Andalucía” within the program 54A “Scientific Research and Innovation” and by ERDF Andalusia 2021-2027 Program, within the Specific Objective RSO1.1 "Developing and improving research and innovation capabilities and assimilating advanced technologies."








\end{document}